\documentclass[12pt]{article}
\usepackage{amsmath}
\usepackage{amssymb}

\newcommand{\m}{\mu}
\newcommand{\n}{\nu}

\renewcommand {\eqref} [1] {(\ref {#1})}
\newcommand {\slsh} [1] {\not{\hbox{\kern-2pt${#1}$}}}

\newcommand {\be} {\begin{equation}}
\newcommand {\ee} {\end{equation}}
\newcommand {\ber}{\begin{eqnarray*}}
\newcommand {\eer} {\end{eqnarray*}}
\newcommand {\bea}{\begin{eqnarray}}
\newcommand {\eea} {\end{eqnarray}}

\newcommand{\ti}{\tilde}
\renewcommand{\th}{\theta}
\newcommand{\D}{\Delta}
\newcommand{\G}{\Gamma}

\def\dd{\mbox{d}}
\newcommand{\sm}[1]{\mbox{\scriptsize #1}}
\def\half{{1\over2}}
\def\f{\phi}
\def\F{\Phi}
\def\vf{\varphi}
\def\pa{\partial}
\def\l{\lambda}
\def\d{\delta}
\def\nn{\nonumber\\}
\newcommand\eq [1] {(\ref {#1})}

\def\bra{\langle}
\def\ket{\rangle}

\newcommand\LHS{left-hand side}
\newcommand\RHS{right-hand side}
\newcommand\thalf{\tfrac12}
\newcommand\iy\infty
\newcommand{\hyp}[5]{\,\mbox{}_{#1}F_{#2}\left(
  \genfrac{}{}{0pt}{}{#3}{#4};#5\right)}
\newcommand{\bisub}[2]{\genfrac{}{}{0pt}{1}{#1}{#2}}

\begin{document}
\begin{flushright}
AEI-2006-046\\
\end{flushright}
\vskip0.1truecm

\begin{center}
\vskip 3truecm {\Large \textbf{Instantons and Conformal Holography}}
\vskip 2truecm

{\large \textbf{Sebastian de Haro${}^\star$ and Anastasios C.~Petkou${}^\dagger$}}\\

\vskip .4truecm
\vskip 4truemm ${}^\star${\it Max-Planck-Institut f\"{u}r Gravitationsphysik\\
Albert-Einstein-Institut\\
14476 Golm, Germany}\\
\tt{sdh@aei.mpg.de}\\
\vskip .3truecm
$^\dagger${\it University of Crete, Department of Physics\\
71003 Heraklion, Greece}\\
{\tt petkou@physics.uoc.gr}

\end{center}
\vskip 1truecm

\begin{center}

\textbf{\large \bf Abstract}

\end{center}

We study a subsector of the AdS$_4$/CFT$_3$ correspondence where a class of solutions in the bulk and on the boundary can be explicitly compared. The bulk gravitational theory contains a conformally coupled scalar field with a $\phi^4$ potential, and is holographically related to 
a massless scalar with a $\phi^6$ interaction in three dimensions. We consider the scalar sector of the bulk theory and match bulk and boundary classical solutions of the equations of motion. Of particular interest is the matching of the bulk and the boundary instanton solutions which underlies the relationship between bulk and boundary vacua with broken  conformal invariance. Using a form of radial quantization  we show that quantum states in the bulk correspond to multiply-occupied single particle quantum states in the boundary theory. This allows us to
explicitly identify the boundary composite operator which is dual to the bulk scalar, at the free theory level  as well as in the instanton vacuum. We conclude with a discussion of possible implications of our results.

\newpage

\tableofcontents

\newpage

\section{Introduction and Motivation}

The deep relationship between quantum field theory and gravity via holography has provided a very strong drive in the current research in string theory. In particular, 
classical field dynamics in AdS spaces has enjoyed a large amount of attention over the last few years and an enormous amount of work on classical supergravities in various dimensions has found important applications in classical gravity, Standard Model phenomenology, and Cosmology.

Progress, however, has been limitied by what is also
one of the virtues of AdS/CFT \cite{maldacena,witten,adsreview}, namely the fact that it is a weak/strong coupling duality. In practice this means that the boundary field theories, being strongly coupled,  are usually not under control. There are very few examples where the holography of controlable quantum field theories can be studied in perturbation theory. Higher-Spin gauge theories\footnote{See \cite{Misha} for a recent review and further references.} might be such an example as it has been argued that they provide the holographic dual of free-field theories. On the other hand, there are examples where the boundary field theory is basically unknown, such as the holographic dual of the M2 brane.\footnote{See \cite{Schwarz} for a recent discussion.} Both examples above involve the AdS$_4$/CFT$_3$ correspondence, which in many respects seems rather special. 

The crucial observation that has sparked our investigation is the existence of classical instanton solutions to the equations of motion both for the bulk and for the boundary theories and that there is an apparent relationship between them: {\it bulk instantons restricted to the boundary are the square of boundary instantons}. Remarkably, this toy model touches upon both Higher-Spin gauge theories, which possess a conformally coupled scalar with a quartic self-interaction, as well as the low-energy action of the M2 brane \cite{hh}. We hope that our results might help shedding more light on the above two important problems. 

One of the aims in this work is to present a simple toy-model of 
the AdS$_4$/CFT$_3$ correspondence where both the bulk and boundary theories are under control. This model, already studied in \cite{hh}, contains a conformally coupled scalar with $\phi^4$ interaction in the bulk of AdS$_4$, which is holographically related to a massless scalar with a $\phi^6$ interaction in three dimensions. Despite the drastic simplification of the boundary theory to a single scalar field, we find that it reproduces the bulk results remarkably well for a large class of solutions. In particular, there are bulk instanton solutions that preserve AdS boundary conditions and correspond to the expectation value of a composite operator of dimension 1 in the boundary theory, which is reproduced by a classical computation in the model CFT. Since we will be interested in this single boundary operator, we restrict to the scalar sector in the bulk. We further simplify the model by neglecting the back-reaction which is partly justified in our linearized approximation since we expand around exact solutions of the equations of motion whose improved stress-energy tensor is identically zero \cite{nextpaper}. A full holographic analysis, for example the discussion of the boundary two-point functions, does require the inclusion of back-reaction of the fluctuations in the case of instanton solutions. In the bulk, such effects can be introduced systematically and these results will be presented in the followup paper \cite{nextpaper}. 

Our model should be viewed as a subsector of the usual AdS$_4$/CFT$_3$ correspondence. Issues such as the counting of the degrees of freedom or the stress-energy tensor of the boundary theory should be analyzed in the coupled scalar-gravity system. The same is true for the boundary theory, where a more refined definition may be needed in order to compute more general observables. We will leave such issues for the future.

Since our goal is not to explore the microscopic boundary theory directly, which as explained is very difficult in the case of AdS$_4$/CFT$_3$, we proceed by constructing an effective theory on the boundary and showing that it reproduces the bulk results in the subsector that is being studied. This theory will be a semi-classical theory, therefore both sides of the duality should be under control. As we will explain, a key feature which makes this possible is the existence of bulk and boundary instanton solutions with a large conformal symmetry group. In order to distinguish this type of holography, where classical solutions can be matched, from the usual one, where a classical theory for the strongly coupled fields is not known, we will denote it conformal holography. Of course, conformal holography is a particular case of classical holography where conformal invariance and the use instanton-like solutions allow us to formulate a classical effective theory for the strongly coupled fields. For the types of models under consideration, this way of doing holography may bring us into a broken phase of the holographic boundary theory.

We now summarize our main set-up and results. We consider the $\f^4$ scalar field theory in the bulk, which
is classically invariant under four-dimensional Euclidean conformal transformations, $O(5,1)$. The way our solutions are constructed is by demanding that they preserve a large symmetry group. It turns out that in the interacting theory the only solution invariant under the full Poincare group is the trivial one. Thus we look for solutions that preserve a subgroup of the symmetry group, namely $O(4,1)$. Notice that the solution is unique once one demands which symmetry one wants it to preserve. Here it is crucial that we are dealing with an interacting theory and not with a free theory, as it is the interactions which determine the solution. The solution is the analytic continuation of the well-known Fubini-Lipatov scalar field solution \cite{fubini,Lipatov}. Such solutions have finite action and are subdominant at weak coupling, whereas at strong coupling they will plague the vacuum.  Although we cannot claim to have full control over interactions, nevertheless one can argue that taking into account other effects like back-reaction shouldn't drastically change our picture.

The next question is whether there is a classical boundary theory whose solutions reproduce the bulk instanton solution. When restricted to the boundary, we still have a large symmetry group, $O(3,1)$, which is the (Euclidean and global) conformal group in two dimensions. Here it  becomes clear that the boundary theory we are looking for is not in its normal vacuum, but in a broken one. It turns out that there is such a theory, namely a massless scalar field theory with a $\phi^6$ interaction. We find it an encouraging sign that we find precisely this theory, which has earlier been advocated as a possible candidate for an effective theory on the boundary \cite{hh,EGPR}. It turns out that also this theory has a unique instanton solution that preserves $O(3,1)$. As mentioned at the beginning, the bulk instanton solution is the square of the boundary instanton.

After having discussed our vacuum state and found agreement between bulk and boundary, we set out to discuss fluctuations. At this point we no longer have exact solutions, but this is irrelevant as we have already expanded around an exact vacuum and all we are interested in is the fluctuations around it. In the interacting case we find once more that the bulk solution is the square of the boundary solution, but it involves a mixing of modes with different conformal dimensions. The reason for this is that the instanton breaks dilatation invariance.

Having the classical correspondence between the bulk and boundary modes, we then quantize them using radial quantization. In this way we are able to identify the composite operator of dimension 1 in the boundary theory as the normal-ordered product of two elementary operators of dimension $1/2$:
\bea
\F_{\sm{hol}}=\,
{}\{\vf(x)^2\}~.
\eea
Is is gratifying to also find that the classical bulk theory is automatically compared to a renormalized operator

Finally, we discuss how the instanton can be seen as a tunneling solution between two local vacua. The fluctuations around the instanton find 
themselves surrounded by a local effective de Sitter geometry where they acquire tachyonic mass. These considerations may add some cosmological interest to our model.

\section{Classical Holography}\label{classicalhol}

In this section we review the conventional holographic picture, where correlation functions of composite CFT operators, and sources for these 
operators, are obtained from classical computations in the bulk of AdS. We will describe the holography of instantons in our toy model, and 
present our proposal for the boundary theory to which they can be explicitly compared.

\subsection{The Model}

Our toy model is a conformally coupled scalar field on a fixed AdS$_4$ background with action
\be\label{I0}
I =\int \dd^4 x\,\sqrt{g}\left(\frac{1}{2}\, g^{\m\n}\partial_\mu\phi\,\partial_\nu\phi -\frac{1}{L^2}\,\phi^2 +\frac{\lambda}{4!}\,\phi^4\right).
\ee
We take the Poincare patch of Euclidean
 AdS$_4$ with metric
 \be
 \label{metricPoin}
 \dd s^2 =g_{\mu\nu}\,dx^\mu dx^\nu=\frac{L^2}{r^2}\left({\over}\!\dd r^2+\dd\vec{x}^2\right)\,,\,\,\,\,x^\mu=(r,\vec{x})\,,\,\,\,\, r>0\,.
 \ee
We rescale the metric and the scalar field as
\be
g_{\mu\nu} =\Omega^{-2}(x)\,\eta_{\mu\nu} \,,\,\,\,\, \phi(x)=\Omega(x)\,\F(x)\,,\,\,\,\,\Omega(x)=\frac{r}{L}\,,
\ee
and the action becomes
\be
\label{I1}
I_\epsilon = \frac{1}{2}\int \dd^3\vec{x}\,\frac{1}{r}\,\F^2(x)\Bigl|^{\infty}_\epsilon +\int_{{\mathbb{R}}^4_{+,\epsilon}}\dd^4x \left(\frac{1}{2}\,
\eta^{\mu\nu}\partial_\mu \F\,\partial_\nu \F+\frac{\lambda}{4!}\,\F^4\right)\,,
\ee
where we have momentarily restricted the holographic variable to $\epsilon \leq r <\infty$. The first term on the right-hand side of (\ref{I1}) is a boundary 
contribution and the second term is the flat space action of a massless $\phi^4$ theory on the half-space ${\mathbb{R}}^4_{+,\epsilon}$ 
with $\epsilon\leq r<\infty$. Notice that for the conformally coupled scalar on a fixed AdS$_4$ background  the only remainder of the bulk 
curvature is the first divergent term. This is not true for any other value of the bulk mass. 

We term Classical Holography the procedure of calculating the on-shell value of the bulk action and its subsequent interpretation as the 
renormalized generating functional of a boundary CFT \cite{witten}. In particular, one aims in obtaining a finite quantity when  
$\epsilon\rightarrow 0$ and to achieve that in our case we subtract the first term on the right-hand side  of (\ref{I1}):
\bea
\label{IR}
I_{\sm{ren}}&\equiv &\lim_{\epsilon\rightarrow0}\Biggl[I_\epsilon-\frac{1}{2}\int \dd^3\vec{x}\,\frac{1}{r}\,\F^2(x)\Bigl|^{\infty}_\epsilon \Biggl]\nonumber \\
&=& \int_{{\mathbb{R}}^4_+}\dd^4x \left(\frac{1}{2}\,\eta^{\mu\nu}\partial_\mu \F\,\partial_\nu \F+\frac{\lambda}{4!}\,\F^4\right)\,.
\eea
For the conformally coupled scalar, holographic renormalization with a single counterterm is 
enough. This would no longer be true for any other value of the bulk mass \cite{SSK}.\footnote{See \cite{Kostas} for a review of 
holographic renormalization.}
Hence, Classical Holography of the conformally coupled scalar is a classical field theory problem in a flat space with a boundary. 
The on-shell value of the action (\ref{IR}) is
\be
\label{IRos}
I_{\sm{ren}}^{\sm{on-shell}} =\frac{1}{2}\int \dd^3\vec{x}\,\F\,\partial_r \F\Bigl|^{\infty}_0-\frac{\lambda}{4!}\,\int \dd^4x\, \F^4\,,
\ee
where we need to solve the  equation of motion
\be
\label{eom1}
\left(\partial_r^2 +\vec{\partial}^2\right)\F(r,\vec{x}) =\frac{\lambda}{3!}\,\F^3(r,\vec{x})\,,
\ee
with Dirichlet boundary conditions at $r=0$. Equation (\ref{eom1}) can be solved perturbatively in $\lambda$ using the standard Green's function for the Dirichlet problem in half space as
\be
\label{solut1}
\F(r,\vec{x}) =\F_0(r,\vec{x}) +\frac{\lambda}{3!}\int \dd^4y\, G(x,y)\,\F^3(y)\,,
\ee
where
\be
\left(\partial_r^2 +\vec{\partial}^2\right)\F_0(r,\vec{x})=0\,,\,\,\,\left(\partial_r^2 +\vec{\partial}^2\right)G(x,y) = \delta^4(x-y) 
\,,\,\,\,\, G(x,y)\Bigl|_{\partial {\mathbb{R}}^4_+} =0\,.
\ee
The asymptotic behavior of $\F_0$ near the boundary is
\be
\label{asympt1}
\F_0(r,\vec{x}) = \alpha(\vec{x}) +r\,\beta(\vec{x}) +\cdots
\ee
with $\alpha$ and $\beta$  arbitrary functions and the dots standing for higher powers in $r$. Requiring that the solution is regular at $r=\infty$ gives the textbook 
formula
\bea
\label{solut11}
\F_0(r,\vec{x}) &=&\frac{1}{\pi^2}\int \dd^3\vec{y}\,\frac{r}{[r^2+(\vec{x}-\vec{y})^2]^2}\,\alpha(\vec{y})\nn
&=&\alpha(\vec{x}) +r\,\frac{1}{\pi^2} 
\int \dd^3\vec{y}\,\frac{1}{(\vec{x}-\vec{y})^4}\,\alpha(\vec{y}) +\cdots\,,
\eea
which determines essentially the relation between $\alpha$ and $\beta$ and yields the boundary two-point function. 
Using the above result we can expand the solution (\ref{solut1}) in powers of $\lambda$ and substitute back into (\ref{IRos}). This gives a 
functional of $\alpha$ which is identified with minus the generating functional for connected correlation functions of a operator ${\cal O}_2$ 
in a three-dimensional CFT that lives on ${\mathbb{R}}^3$
\bea
\label{IRos1}
-I_{\sm{ren}}^{\sm{on-shell}}[\alpha] &=& W_{\sm{ren}}[\alpha] =\frac{1}{2}\int \dd^3\vec{x}\,\alpha(\vec{x})\,\beta(\vec{x}) +O(\lambda)\,,\\
\label{W}
e^{W_{\sm{ren}}[\alpha]}&=& \int [{\cal D}\phi]\,e^{-S[\phi] +S_{\mbox{\tiny{int}}}[\alpha, {\cal O}_2]}\,,\,\,\,\, S_{\sm{int}}[\alpha,{\cal O}_2] =\int \dd^3\vec{x}\,\alpha\,{\cal O}_2\,.
\eea
In most examples of holography the boundary CFT is strongly coupled and the action $S[\phi]$ is unknown. Then, one finds from (\ref{IRos1}) using (\ref{solut1}) 
\bea
\label{NV2pt1}
\frac{\delta W_{\sm{ren}}[\alpha]}{\delta\alpha(x)} =\beta(\vec{x})\equiv \langle {\cal O}_2(x)\rangle_{\alpha} &=&\int \dd^3\vec{x}_1\,\langle {\cal O}_2(\vec{x}){\cal O}_2(\vec{x}_1)\rangle\, \alpha(\vec{x}_1) +O(\lambda)\,,\nn
\langle{\cal O}_2(\vec{x}_1){\cal O}_2(\vec{x}_2)\rangle &=&\frac{1}{\pi^2(\vec{x}_1-\vec{x}_2)^4}\,.
\eea
The above show that the scalar operator ${\cal O}_2$ has dimension $\Delta=2$.

Choosing  $\alpha$ as the boundary source is not the only possibility for the conformally coupled scalar. We could have chosen to express the boundary on-shell action in terms of $\beta$ and still get a consistent generating functional. The correct way to do that is to consider the Legendre transform of $W_{\sm{ren}}[\alpha]$ as \cite{Kleb_Witten}
\be
\label{LT1}
W_{\sm{ren}}[\alpha] =\Gamma[{\cal A}] +\int \alpha{\cal A}\,,\,\,\,\,\,\frac{\delta W[\alpha]}{\delta\alpha(x)} ={\cal A}(x) \,,\,\,\,\,\,\frac{\delta\Gamma[{\cal A}]}{\delta{\cal A}(x)} =-\alpha(x)\,.
\ee 
Setting $A(\vec{x})=\beta(\vec{x})$ this will give the generating functional of the Dual boundary CFT as
\be
\label{Legendre1}
\Gamma[\beta]\equiv \tilde{W}_{\sm{ren}}[\beta] =\frac{1}{2}\int \dd^3\vec{x}\dd^3\vec{y}\, \frac{C_1}{(\vec{x}-\vec{y})^2}\,\beta(\vec{x})\,\beta(\vec{y})
+O(\lambda)\,,\,\,\,\,\, C_1=\frac{1}{2\pi^2}\,.
\ee
The 1- and 2-pt functions of the Dual boundary CFT are
\bea
\label{NV2pt1_2}
\frac{\delta \tilde{W}_{\sm{ren}}[\beta]}{\delta\beta(x)} \equiv -\alpha(\vec{x}) =\langle {\cal O}_1(x)\rangle_{\beta} &=&\int \dd^3\vec{x}_1\langle{\cal O}_1(\vec{x})\,{\cal O}_1(\vec{x}_1)\rangle\beta(\vec{x}_1) +O(\lambda)\,,\nn
\label{2-pthol}
\langle{\cal O}_1(\vec{x}_1)\,{\cal O}_1(\vec{x}_2)\rangle &=&\frac{1}{2\pi^2(\vec{x}_1-\vec{x}_2)^2}~.
\eea
Therefore, the Dual boundary CFT has a scalar operator ${\cal O}_1$ with dimension $\Delta=1$. This is the operator we will consider in this paper and we will denote it hereafter  by ${\cal O}\equiv {\cal O}_1$.

\subsection{Instantons and the Vacuum Structure of the Boundary CFTs}

The vacuum structure of the boundary CFTs is probed by the external sources. For example, setting $\alpha(\vec{x})=0$ we find that the standard boundary CFT is in its normal vacuum where $\langle {\cal O}_2\rangle =0$. Equivalently, setting $\beta(\vec{x})=$ we see that $\langle {\cal O}\rangle =0$ for the Dual boundary CFT as well. Two other interesting boundary configurations are
\bea
{\rm Standard\,\,\,CFT:}\,\,\,\,\, \alpha(\vec{x})=\delta^3(\vec{x}) &\Rightarrow& \beta(\vec{x}) \equiv \frac{C_2}{\vec{x}^4} =\langle{\cal O}_2(\vec{x}){\cal O}_2(0)\rangle\\
{\rm Dual\,\,\,CFT:}\,\,\,\,\, \beta(\vec{x}) =\delta^3(\vec{x}) &\Rightarrow & 
\alpha(\vec{x}) =-\frac{C_1}{\vec{x}^2} =-\langle{\cal O}(\vec{x})\,{\cal O}(0)\rangle
\eea
Notice that a delta function source for ${\cal O}$ in the Dual boundary CFT corresponds to negative values of the field at the boundary.

The Legendre transform functional $\Gamma[\beta]$, being an effective potential for the standard boundary CFT, determines its vacuum structure.  For example, in the absence of external sources for ${\cal O}_2$ we have 
\be
\label{A0}
\langle {\cal O}_2(\vec{x})\rangle_{\alpha=0} =\alpha_0(\vec{x})\,,\,\,\,\,\,\,\,\frac{\delta\Gamma[{\cal A}]}{\delta\alpha(\vec{x})}\Bigl|_{\alpha_0}=0\,.
\ee
For $\alpha_0=0$ we have the normal vacuum, for $\alpha_0\neq 0$ we have a broken vacuum. In the latter case, the expectation value of ${\cal O}_2(\vec{x})$ is non-zero and in translationally invariant theories it is a constant. 

In the Dual boundary CFT the roles of $\Gamma[\beta]$ and $W_{\sm{ren}}[\alpha]$ are interchanged. Now the latter becomes the effective potential as is seen from 
\be
\label{LT2}
\frac{\delta\Gamma[\beta]}{\delta\beta(\vec{x})} =-\alpha(\vec{x})=\langle {\cal O}(\vec{x})\rangle_\beta\,,\,\,\,\,\,\,\frac{\delta W_{\sm{ren}}[\alpha]}{\delta\alpha(\vec{x})} =\beta(\vec{x})
\ee
Then, $\beta=0$ is the extremization condition for the effective potential of the Dual boundary CFT and determines its vacuum structure. 
\be
\label{NewVac}
\langle{\cal O}(\vec{x})\rangle_{\alpha=0} =-\alpha_0(\vec{x}) \,,\,\,\,\,\,\,\frac{\delta W[\alpha]}{\delta\alpha(\vec{x})}\Bigl|_{\alpha_0}=0~.
\ee
Hence, {\it the vacuum structure of the Dual boundary CFT can be found by extremizing $W_{\sm{ren}}[\alpha]$.} 

The above discussion finds its application in our toy model. 
When $\lambda<0$, a real solution of (\ref{eom1}) exists and is given by the Fubini-Lipatov instanton \cite{fubini,Lipatov}
\be
\label{FubLip}
\F_0(\vec{x}) =\sqrt{\frac{48}{-\lambda}}`\,\frac{b}{b^2 +r^2 +\vec{x}^2}\,.
\ee
This depends on an arbitrary parameter $b$ with dimensions of length, the instanton size. The case $\l>0$ will be discussed in section 5. The instanton action is independent of $b$ and is evaluated to 
\be\label{I00}
I_0 =-\frac{8\pi^2}{\lambda}\,.
\ee 
We expand around the solution (\ref{FubLip}) as 
\be
\F(\vec{x}) =\F_0(\vec{x}) +\tilde{\F}(\vec{x}) \,,\ee
and obtain
\be
\label{Iinst}
I_{\sm{ren}} = I_0 +\int \dd^3\vec{x}\, \tilde{\F}\partial_r \F_0\Bigl|_0^\infty +\int_{R^4_+}\dd^4x \left(\frac{1}{2}\,\eta^{\mu\nu}\partial_\mu\,\tilde{\F}\partial_\nu\tilde{\F} -\frac{12b^2}{(b^2+r^2+\vec{x}^2)^2}\,\tilde{\F}^2 +V(\tilde{\F})\right)\,,
\ee
where the potential $V(\tilde{\F})$ contains cubic and higher terms in $\tilde{\F}$.  
To calculate the on-shell action we need to solve the equation of motion for the fluctuations $\tilde{\F}$ . We restrict ourselves to the linearized fluctuation equations 
\be
\label{lineom}
\left(\partial^2_r+\vec{\partial}^2 +\frac{24b^2}{(b^2+r^2+\vec{x}^2)^2}\right)\tilde{\F}(r,\vec{x})=0\,.
\ee
The general solution of (\ref{lineom}) behaves near $r=0$ as
\be
\label{r0tildef}
\tilde{\F}(r,\vec{x}) \approx \tilde{\F}_0(\vec{x}) +r\,\tilde{\F}_1(\vec{x}) +\cdots\,.
\ee
In Classical Holography we should solve (\ref{lineom}) imposing Dirichlet boundary conditions at $r=0$ and regularity at $r=\infty$ as
\be
\tilde{\F}(0,\vec{x}) ={\F}_0(\vec{x})\,,\,\,\,\,\,\tilde{\F}\left(r=\infty,\vec{x}\right) =0\,.
\ee
This is done in Appendix \ref{standardsol}.
Then, the quadratic on-shell action as a functional of the boundary conditions is
\be
\label{IRosa}
-I_{\sm{ren}}^{\sm{on-shell}}[\alpha] \equiv W_{\sm{ren}}[\alpha] =\frac{8\pi^2}{\lambda} +\frac{1}{2}\int \dd^3 x\,\tilde{\F}(\vec{x})\,\tilde{\F}_1(\vec{x})
\ee
where we have denoted
\be
\label{alpha}
\alpha(\vec{x}) =\F_0(\vec{x}) +\tilde{\F}(\vec{x})~.
\ee
Clearly, $\delta\alpha =\delta\tilde{\F}$ and we find
\be
\label{dWR}
\frac{\delta W_{\sm{ren}}[\alpha]}{\delta\alpha(\vec{x})} =\frac{1}{2}\,\tilde{\F}_1(\vec{x}) +\frac{1}{2}\int \dd^3y\, \tilde{\F}(\vec{y})\,\frac{\delta\tilde{\F}_1(\vec{y})}{\delta\tilde{\F}(\vec{x})}
\ee
In Appendix \ref{standardsol} we show that setting $\tilde{\F}(\vec{x})=0$ also gives $\tilde{\F}_1(\vec{x})=0$, therefore the boundary effective action is minimized (recall that $\lambda<0$),  when 
\be
\alpha(\vec{x})=\alpha_0(\vec{x}) =\F_0(\vec{x})\,.
\ee
We define the Legendre transform functional $\Gamma[{\cal A}]$ as
\be
\label{LegendreTr}
W_{\sm{ren}}[\alpha]=\Gamma[{\cal A}] +\int \dd^3x\,(\tilde{\F} +\alpha_0){\cal A}\,,\,\,\,\frac{\delta W_{\sm{ren}}[\alpha]}{\delta\alpha} ={\cal A} \,,\,\,\,\,\frac{\delta \Gamma[{\cal A}]}{\delta{\cal A}} =-(\alpha+\alpha_0)\,,
\ee
and we interpret it as the generating functional for composite operators coupled to the source ${\cal A}$ in the Dual boundary CFT.  Then we see that the theory described by $\Gamma[{\cal A}]$ has a vacuum state in which the operator coupled to ${\cal A}$ gets a non-zero expectation value, after having switched off the external source, given by
\be
\label{1ptO1}
\frac{\delta \Gamma[{\cal A}]}{\delta\alpha(\vec{x})}\Bigl|_{{\cal A}=0} \equiv \langle {\cal O}(\vec{x})\rangle_{{\cal A}=0} =-\alpha_0(\vec{x}) =-\sqrt{\frac{48}{-\lambda}}\,\frac{b}{b^2+\vec{x}^2}\,.
\ee

\subsection{The Proposal for the Dual boundary CFT}

The Dual boundary CFT of our toy model is a CFT which has a scalar operator with dimension $\Delta=1$. A natural candidate for this is provided by a massless scalar in three dimensions. Indeed, the elementary scalar field $\vf$ (i.e. the one appearing in the Lagrangian) in three dimensions has dimension $\Delta=1/2$, hence one might consider the composite operator $\vf^2$ as a candidate for the operator ${\cal O}$.  Clearly, this is is not a very strong argument since as we discussed previously in Classical Holography ${\cal O}(\vec{x})$ is just an arbitrary function. Nevertheless, instantons provide more solid evidence for such correspondence. In this case, the value of ${\cal O}$ is fixed at the boundary to be (\ref{1ptO1}). Can this value be related to the value of $\vf^2$ in some boundary theory? The answer is, remarkably, positive. 

Consider the massless scalar in three dimensions with a $\vf^6$ interaction
\be\label{bdyaction}
S=\int\dd^3x\,[\half(\pa_\m\vf)^2 +{g\over6!}\,\vf^6]~.
\ee
When the dimensionless coupling constant $g<0$, the equations of motion have the real instanton solution \cite{fubini}
\be
\label{Inst3}
\vf_0(\vec{x}) =\left(\frac{360}{-g}\right)^{1/4}\left(\frac{c}{c^2+\vec{x}^2}\right)^{1/2}\,,
\ee
where the instanton size $c$ is an arbitraty parameter with dimensions of length. These instantons have finite action, independent of $c$,  which is
\be
\label{IAction3}
S_0=\frac{3\pi^2}{2}\sqrt{\frac{10}{-g}}\,.
\ee
The results (\ref{1ptO1}) and (\ref{Inst3}) are consistent with the  correspondence between ${\cal O}$ and $\vf^2$. Since the bulk Hilbert space is a product of two boundary Hilbert spaces one expects that the actions (\ref{I00}) and (\ref{IAction3}) are related as
\be
\label{gl}
I_0=2S_0\Rightarrow \frac{1}{g} =-\frac{32}{45}\frac{1}{\lambda^2}\,.
\ee
Then we can identify the bulk instantons with the square of the boundary instantons after a rescaling by a dimensionless paramater $\kappa$ of their size and position as
\be
c=\kappa\, b\,,\,\,\,\,\, \vf_0^2(\kappa\,\vec{x}) =-\langle {\cal O}(\vec{x})\rangle \Rightarrow \kappa^2=\frac{16\pi^2}{3\lambda}\,.
\ee
Therefore, qualitatively our proposal is that the bulk model (\ref{I0}) with negative values for $\lambda$ is holographically dual to the model (\ref{bdyaction}) with negative coupling constant given by (\ref{gl}). 
We should remark here that the holographic comparison of the finite part of the action is subject to the usual renormalization scheme ambiguities. In the next few sections we will add evidence to the above proposal.

For positive values of $g$, the theory (\ref{bdyaction}) is only well defined in the context of some large-$N$ expansion \cite{bardeen}. We expect that the explicit form of the bulk/boundary correspondence of our model (\ref{I0}) will involve some large-$N$ expansion which, however, we shall not try to  identify here.  

\section{Conformal Holography: Free Case}

In the previous section we reviewed the standard holographic procedure, termed Classical Holography, in which the on-shell bulk action is interpreted as the renormalized generating functional of the boundary theory. In this section we show how in our toy model we can go one step further and, after we match bulk and boundary classical field configurations, upon quantization we  can identify quantum modes in the bulk with quantum modes on the boundary. This way we reproduce the results of  section 2 for the expectation value of the operator of dimension 1, but we also find a microscopic description of it in the bulk. In this section we restrict ourselves to the free case, i.e.~$\l=0$.

\subsection{Geometric Set-up for  Radial Quantization}\label{geomsetup}

Four-dimensional Euclidean anti-de Sitter space is a hyperboloid in 5-dimensional Minkowski space with metric $\eta_{AB}=\mbox{diag}(-1,1,1,1,1)$ specified by the constraint
\be\label{hyperboloid}
\eta_{AB}\,y^A y^B=-y_0^2+y_4^2+\vec{y}^2=-L^2~,
\ee
where $y^A=(y_0,\vec{y},y_4)$ and  $\vec{y}=(y_1,y_2,y_3)$. Its isometry is group $O(4,1)$. 
Notice that this space has two disconnected 
branches, $y_0\geq L$ and $y_0\leq-L$.  

We solve the above constraint by introducing the following set of global coordinates
\bea\label{cylindercoords}
u&=&y_0+y_4={R\over\cos\th}\,,\nn
v&=&y_0-y_4={L^2\over R\cos\th}\,,\nn
\vec{y}&=&L\tan\th\,\vec\Omega_2\,,
\eea
where $\vec\Omega_2$ as usual parametrizes the unit 2-sphere, with volume element $\dd\Omega_2^2=\dd\psi^2+\sin^2\psi\,\dd\omega^2$. In these coordinates,
the AdS$_4$ metric takes the form
\be\label{cylinder}
\dd s^2={L^2\over R^2\cos^2\th}\left(\dd R^2+R^2\dd\Omega_3^2\right)~,
\ee
where $\dd\Omega_3^2=\dd\th^2+\sin^2\th\,\dd\Omega_2^2$, and $0\leq \theta\leq \frac{\pi}{2}$. Defining $R/L=e^{\tau/L}$ we also obtain
\be
\label{cylinder2}
\dd s^2 =\frac{1}{\cos^2\th}\left(\dd\tau^2+L^2\,\dd\Omega^2_3\right)~.
\ee
The above is seen to be conformal to a ``half-cylinder''  ${\mathbb{R}}\times S^3_+$ with $S^3_+$ being half a 3-sphere. The latter space has a boundary at $\th=\pi/2$, which is the conformal boundary of AdS$_4$ . 

Unlike the Lorentzian case where $\tau$ is a global coordinate, we see that here $\tau\in(-\infty,\infty) $ only covers the region $R>0$.
Whereas the boundary of Lorentzian AdS$_4$ is $(S^1\times S^2)/{\mathbb{Z}}_2$, in the Euclidean case the boundary is non-compact and topologically
$({\mathbb{R}}\times S^2)/{\mathbb{Z}}_2$. This is easy to see; sending $y_A\rightarrow\infty$ (equivalently,  $\th\rightarrow\pi/2$)
while preserving \eq{hyperboloid}, we get the constraint
\be\label{bdydef}
\tilde u\tilde v=\vec{\tilde{y}}^2=1~,
\ee
where $\vec{\ti y}$ are now coordinates on the boundary, obtained by a proper rescaling of the bulk coordinates. We can solve this as
\bea\label{bdycoords}
\ti y_0&=&\cosh\tau\nn
\ti y_4&=&\sinh\tau\nn
\vec{\ti y}&=&\vec\Omega_2~.
\eea
$\tau$ is related to $R$ as before. $(R,\psi,\omega)$ parametrize ${\mathbb{R}}\times S^2$ on the boundary. However, they do not solve the constraint \eq{bdydef} completely. We
sill have to divide out the ${\mathbb{Z}}_2$ action $\ti y_A\sim-\ti y_A$, which is $R\sim -R$, $\vec\Omega_2\sim-\vec\Omega_2$. Thus, the 
boundary is topologically $({\mathbb{R}}\times S^2)/{\mathbb{Z}}_2$. 
For later convenience, we list here the two $O(4,1)$ discrete symmetries that we will use in this paper:
\bea\label{discretesym}
T &:&y_4\rightarrow-y_4~,~{R\over L}\rightarrow{L\over R}\nn
P &:&y_0\rightarrow-y_0~,~R\rightarrow-R \,\,\,~~{\rm and}~~\,\,\,\vec y\rightarrow-\vec y~\,,~\vec\Omega_2\rightarrow-\vec\Omega_2
\eea
$T$ is the time reversal operation in Euclidean time $\tau$, and generates inversions in $R$. It will be an important ingredient of radial
quantization. $P$ is the parity operation, which takes a point $x=(R,\th,\vec\Omega_2)$ to its antipode $Px=(-R,\th,-\vec\Omega_2)$. When it acts on
$y_0$, it interchanges the two disconnected branches of AdS$_4$. 
The boundary of AdS$_4$, as obtained above, is ${\mathbb{R}}\times S^2$ 
moded out by the action of $P$. 








\subsection{Classical Correspondence}

That a conformally coupled scalar $\phi$ in AdS$_4$ is related to a massless scalar in three dimensions can already be seen at the classical level. Considering for simplicity the free case, using the metric (\ref{cylinder}) we find that the equation of motion for the rescaled bulk scalar $
\Phi(R,\Omega_3)=\phi(R,\Omega_3)/R\cos\th\,$ is just  Laplace's equation in radial coordinates
\be
\Box\Phi=\left(\pa_R^2+{3\over R}\,\pa_R+{1\over R^2}\,\Delta_{S_{\tiny{+}}^3}\right)\Phi=0~.
\ee
Had we considered the full sphere $S^3$, where $0\leq \theta\leq\pi$ the general solution of the above would be  given in terms of the hyperspherical harmonics ${\cal Y}_{jl m}$ (see Appendix A) as
\be\label{PhiLaplace}
\Phi(R,\th,\psi,\vf)=\sum_{j=0}^\infty\sum_{l=0}^j\sum_{m=-l}^l \left(c_{(1)jlm}\,R^j+c_{(2)jlm}\,{1\over R^{j+2}}\right)\,{\cal Y}_{jlm}(\th,\psi,\vf)~,\ee
where $c_{(1)}$ and $c_{(2)}$ are arbitrary coefficients. 
However, on $S_+^3$ we have $0\leq \theta\leq \pi/2$ and the hyperspherical harmonics with {\it either} $j+l$ even {\it or} $j+l$ odd, separately form complete bases. This means that the general solution is still given by linear combination of terms of the form  (\ref{PhiLaplace}) with even and odd $j+l$. This one-parameter family of solutions corresponds to the freedom to choose the boundary conditions at   $\theta=\pi/2$ as we discuss below.

Notice that the equation of motion for the field $\Phi$ is invariant under $O(5,1)$. The subgroup that preserves the background is of course only 
$O(4,1)$. Moroever, the action of the $P,T,$ operations on the classical field is determined  by the coefficients $c_{(1)}$ and $c_{(2)}$. 
For example, $PT$ transforms a mode $R^j{\cal Y}_{jlm}$ into $(-)^{j+l}R^{-j}{\cal Y}_{jlm}$. In radial quantization these operations are connected to Hermitian conjugation and we already see that  $j+l$ even or odd would correspond to two different quantization schemes.

Classically the difference between the $j+l$ even and odd emerges when one considers the behavior of the field $\Phi$ near the boundary at $\theta=\pi/2$ which is 
\be
\Phi(R,\th,\psi,\vf)\sim \Phi_{(0)}(R,\psi,\vf)+(\cos\th)\,\Phi_{(1)}(R,\psi,\vf)+\cdots
\ee
This is determined by the ${\cal Y}_{jlm}$'s. In particular, as $\theta\rightarrow \pi/2$ for $j+l$ even the hyperspherical harmonics reduce to the standard 
two-dimensional spherical harmonics
\be\label{jlmreduction}
{\cal Y}_{jlm}\left(\th={\pi\over2},\psi,\vf\right)=a_{jl}\,Y_{lm}(\psi,\vf)~.
\ee
The constants $a_{jl}$ and some more details on the hyperspherical harmonics are given in formula \eq{ajl} in Appendix \ref{hypersphericalapp}. For $j+l$ odd, the
hyperspherical harmonics go to zero as $\cos\th$. Of course, this discussion simply corresponds to the two quantization schemes discussed
in \cite{ais,bf}, and agrees with the general analysis in \cite{vijayper}. Restricting $j+l$ to be even or odd amounts to making the walls reflective or transparent respectively. Hence, for  $j+l$ even we have an operator with dimension 1 on the boundary, while for  $j+l$ odd the boundary operator has dimension 2. For general integer values of $j+l$ we have both an operator and a source term on the boundary. 

Consider the case when $j+l$ is even. At $\theta=\pi/2$ we get
\be\label{holPhi}
\Phi^{\sm{hol}}(R,\Omega_2)= \sum_{j=0}^\infty\Phi^{\sm{hol}}_j(R,\Omega_2)=\sum_{j=0}^\infty\sum_{l=0}^j\sum_{m=-l}^l\left({c}_{(1)jlm}\,R^j+{c}_{(2)jl m}\,{1\over R^{j+2}}\right)
Y_{lm}(\Omega_2)~,
\ee
where we have reabsorbed the constants $a_{jl}$ in the definition of the coefficients ${c}_{(0)}$ and ${c}_{(1)}$. Since the theory is scale invariant, $R$ can be chosen
to be dimensionless.
Notice that the above scalings $\Delta_+=j+2$, $\Delta_-=-j$ follow from the holographic relationship 
$\Delta_{\pm}={d-1\over2}\pm\sqrt{{(d-1)^2\over4}+m^2}$ applied for a set of scalar fields in $d=3$ with masses $m^2=j(j+2)$. The formula suggests that these scalar fields are dual to operators in a {\it two-dimensional} CFT with dimensions $\D_+=j+2$.

We will now show that this agrees with the classical solutions of the equations of motion of the three-dimensional boundary action \eq{bdyaction}
\be
\Box\vf=\pa_R^2\vf+{2\over R}\,\pa_R\vf+{1\over R^2}\,\Delta_{S^2}\vf=0~.
\ee
Expanding $\vf$ in spherical harmonics we find the general solution of the above
\bea\label{bdysolution}
\vf(R,\Omega_2)&=&\sum_{\ell=0}^\infty\sum_{m=-\ell}^{\ell}\left(\vf_{(1)\ell m}\,R^\ell+\vf_{(2)\ell m}\,{1\over R^{\ell+1}}\right)Y_{\ell m}(\Omega_2)~,\nn
&=&\sum_{\ell=0}^\infty\left(\vf_\ell^+(R,\Omega_2)+\vf_\ell^-(R,\Omega_2)\right)\,,
\eea
where $\vf_{(1)}$ and $\vf_{(2)}$ are arbitrary coefficients. It is then straightforward to establish the classical relation
\be\label{squarerelation}
\Phi^{\sm{hol}}_{j}(R,\Omega_2)= \left(\vf_\ell^+(R,\Omega_2)\right)^2+\left(\vf_\ell^-(R,\Omega_2)\right)^2\,,\,\,\,\,~~ j\equiv 2\ell\,.
\ee
This is done using standard properties of spherical harmonics summarized in the Appendix \ref{hypersphericalapp} as
\bea
\left(\vf_\ell^+(R,\Omega_2)\right)^2&=&R^{2\ell}\sum_{m_1m_2}\vf_{(1)\ell m_1}\vf_{(1)\ell m_2}\sum_{l=0}^{2\ell}\sum_{m=-l}^l c^{\ell\ell m_1m_2}_{l m}\,Y_{lm}(\Omega_2)~;\nn
\left(\vf_\ell^-(R,\Omega_2)\right)^2&=&{1\over R^{2\ell+2}}\sum_{m_1m_2}\vf_{(2)\ell m_1}\vf_{(2)\ell m_2}\sum_{l=0}^{2\ell}\sum_{m=-l}^l c^{\ell\ell m_1m_2}_{lm}\,Y_{lm}(\Omega_2)~.
\eea
Matching this to the holographic bulk field \eq{holPhi} we find
\bea\label{coeffs}
c_{(1)jlm}&=&\sum_{m_1,m_2=-l}^lc_{lm}^{\ell\ell m_1m_2}\,\vf_{(1)\ell m_1}\vf_{(1)\ell m_2}\nn
c_{(2)jlm}&=&\sum_{m_1,m_2=-l}^lc_{lm}^{\ell\ell m_1m_2}\,\vf_{(2)\ell m_1}\vf_{(2)\ell m_2}~,
\eea
with $j=2\ell$. Notice that the condition 
$\ell+\ell'+l={\mbox{even}}$ satisfied by the nonzero coefficients 
$c^{\ell\ell'm_1m_2}_{lm}$ given in the appendix is exactly the condition $j+l={\mbox{even}}$ that we have in the bulk for the theory with an
operator of dimension 1. 
Upon quantization, \eq{coeffs} will become an operator relation between creation and annihilation operators in the bulk and on the boundary and will determine the  relationship between the bulk and boundary Hilbert spaces. 

A priori nothing stops us from comparing the full bulk and boundary fields, instead of comparing single modes as in \eq{squarerelation}. So let 
us try to see whether we can compare $\F_{\sm{hol}}$ with $\vf^2$, including terms with $\ell\not=\ell'$. For example, we can expand
\be\label{q1}
\vf_\ell^+\vf_{\ell'}^+=R^{\ell+\ell'}\sum_{l=|\ell-\ell'|}^{\ell+\ell'}\sum_{m=-l}^l\vf_{(1)\ell m_1} \vf_{(1)\ell' m_2}
\,c^{\ell\ell'm_1m_2}_{lm}\,Y_{lm}
\ee
Comparing this to $\F_{\sm{hol}}^+$, we have to turn the sum over $j$ into a sum over $\ell,\ell'$. This is easy to do:
\bea\label{q2}
\F_{\sm{hol}}^+&=&\sum_{j=0}^\infty\sum_{l=0}^j\sum_{m=-l}^lc_{(1)jlm}\,R^j\,Y_{lm}\nn
&=&\sum_{\ell,\ell'=0}^\infty\sum_{l=0}^{\ell+\ell'}\sum_{m=-l}^l{1\over\ell+\ell'+1}\,c_{(1)jlm}\,R^{\ell+\ell'}\,Y_{lm}~.
\eea
The factor of $\ell+\ell'+1$ in the denominator is simply the multiplicity of terms contributing to $\ell+\ell'=j$. In our earlier formula \eq{coeffs} there was no such factor -- indeed, if we restrict ourselves to the sector $\ell=\ell'$ from the outset (and we will do this in the rest of the paper), there is just one term contributing at each $j$.

There is an important difference when we compare \eq{q1} and \eq{q2}. In the latter, there are contributions from terms with $0\leq l\leq|\ell-\ell'|$, whereas these are absent from \eq{q1}. 

The missing terms notwithstanding, we can still compare both sides for the individual modes. We then get:
\be\label{cjlm}
c_{(1)jlm}=(\ell+\ell'+1)\sum_{m_1m_2}c^{\ell\ell'm_1m_2}_{lm}\,\vf_{(1)\ell m_1}\vf_{(1)\ell' m_2}~,
\ee
where now $j=\ell+\ell'$. In fact, this equation makes sense for any $\ell,\ell'$! The point is that the left-hand side depends only on 
$j=\ell+\ell'$ and not on their difference, but in fact so does the right-hand side. Indeed, using the definition of the coefficients 
$c^{\ell\ell'm_1m_2}_{lm}$ given in Appendix \ref{hypersphericalapp} (see formula \eq{coeffs2}), the right-hand side is easily seen to be
symmetric in $\ell,\ell'$, hence it is only a function of $\ell+\ell'$. Thus at the level of single modes, we can get mixed modes from the 
bulk using \eq{cjlm}. Quantum mechanically, equation \eq{cjlm} becomes an operator relation but we  leave for the future its further exploration.

\subsection{Radial Quantization in the Bulk}

Having at hand the general solution (\ref{PhiLaplace}) we may proceed with its quantization. The expansion in (\ref{PhiLaplace}) is in eigenmodes 
of the dilatation operator $\hat{D}=R(\partial/\partial R)$, hence it is natural to use radial quantization \cite{FHJ}. The arbitrary 
coefficients appearing in (\ref{PhiLaplace}) become creation and annihilation operators. Hermitian conjugation is the earlier discussed $PT$
operation. Since we are in the sector where $j+l$ is even, we simply have \cite{FHJ}
\be
\label{HermConj}
[ \F(R,\Omega_3)]^\dagger = \frac{1}{R^2}\,\F\left(\!{1\over R},\Omega_3\!\right)
\ee
In particular, this ensures that the field is real when we go back to Lorentzian signature.
Without loss of generality we require that the creation operators multiply the modes that are regular at $R=0$. The anihilation operators multiply the modes that are regular at $R=\infty$. Loosely speaking, ``in" states are created at $R=0$ and ``out" states at $R=\infty$. Finally, the Hermitian field operator takes the form
\be
\label{radial_quant}
\hat\F(R,\theta,\psi,\omega)=\sum_{jlm}\left(\frac{a^{+}_{jlm}}{\sqrt{j+1}}\,R^j\, {\cal Y}^*_{jlm}(\Omega_3)+\frac{a^{-}_{jlm}}{\sqrt{j+1}}\,\frac{1}{R^{j+2}}{\cal Y}_{jlm}(\Omega_3)\right)~,
\ee
where $j+l$ is even.

The operators $a^{(+)}_{jlm}$ and $a^{(-)}_{jlm}$ satisfy 
\be
[a^{+}_{jlm}]^{\dagger} =a^{-}_{jlm}\,,\,\,\,\,\,\,\, 
a^{-}_{jlm} |0\ket=0\,,\,\,\,\, \bra0|a^{(+)}_{jlm} =0\,,\,\,\,\,\,\,\,[a^{-}_{jlm},a^{+}_{j'l'm'}] =\delta_{jj'}\delta_{ll'}\delta_{mm'}
\ee
while all other commutators vanish. 

Now, our general holographic interpretation combined with the radial quantization allows us to take the relationship (\ref{NV2pt1_2}) one step further and identify the operator (\ref{radial_quant}), when $\theta=\pi/2$,  with minus the composite operator ${\cal O}$ of dimension 1 in the Dual boundary CFT  
\be
\hat\F\left(R,\frac{\pi}{2},\psi,\omega\right) \equiv -{\cal O}(\vec{x})
\ee
Using (\ref{radial_quant}) we can then calculate the boundary one- and two-point functions of the operator ${\cal O}$. Choosing for simplicity $R> R'$ and the angles such that $\omega=\psi'=\omega'=0$, we obtain 
\be
\bra n|{\cal O}|n\ket=0
\ee
for all multiparticle states $|n\ket$. This is an important property which we should maintain when we try to construct ${\cal O}$ using boundary operators. We also get
\bea
\label{2-ptRQ1}
\bra0|{\cal O}(R,\Omega_2)\,{\cal O}(R',\Omega_2') |0\ket &=&
 \sum_{j=0}^{\infty}\sum_{l=0}^j N_{jl}^2 \left(\frac{\Gamma(l+j+2)}{\Gamma(j-l+1)\Gamma(2l+2)}\right)^2\frac{2l+1}{4\pi} \frac{P_l(\cos\psi)}{j+1}\frac{R'^j}{R^{j+2}}\nonumber \\
  &=& \frac{1}{2\pi^2}\sum_{j=0}^{\infty}C_j^1(\cos\psi)\frac{R'^j}{R^{j+2}} =\frac{1}{2\pi^2}\frac{1}{(\vec{x}-\vec{x}')^2}
\eea
 This is a non-trivial result. The sum in the rhs of the first line in \label{2-ptRQ}) is split into two sums, each one involving terms with {\it either} $j,l$ both being even {\it or} $j,l$ both being odd.  These sums are done using the ``summation theorems" over even and odd integers for Gegenbauer polynomials \cite{gradshteyn}. (\ref{2-ptRQ1}) coincides with the holographic result (\ref{2-pthol}).

In the above computation, summing over all integer $j$'s gives the expected holographic result. In the next section we will match the bulk and boundary Hilbert spaces for even $j$'s. The odd-$j$ sector can be included as well, but the analysis is more involved and we will not do this in this paper. If one restricts the above sum to the even $j$'s, one gets instead:
\be
\bra0|{\cal O}(\vec x)\,{\cal O}(\vec x')|0\ket={1\over 4\pi^2}{1\over(\vec x-\vec x')^2}+{1\over4\pi^2}{1\over(\vec x+\vec x')^2}~,
\ee
that is we get a sum of two images, which now diverge not only at coincident points but also at antipodal points on the boundary. 

\subsection{Mapping of the Hilbert Spaces}

Having defined the quantized bulk field $\Phi$ and seen that it reproduces the 2-point function of the composite operator ${\cal O}(x)$ we ask next whether we can identify this operator explicitly in the boundary theory. The previous discussion points towards a relationship of the form  ${\cal O}(x)\sim \vf(x)^2,$, where $\vf$ is the boundary elementary scalar. 
We radially quantize the boundary field \eq{bdysolution} by promoting the coefficients to operators, as before
\be
\vf(R,\Omega_2)=\sum_{\ell m}{1\over\sqrt{2\ell+1}}\left(b_{\ell m}^\dagger\, R^\ell\, Y_{\ell m}^*(\Omega_2)+b_{\ell m}\,{1\over R^{\ell+1}}\,
Y_{\ell m}(\Omega_2)\right)~.
\ee
Complex conjugation is defined as before, but with weight one: $\f(R)^\dagger={1\over R}\,\f(1/R)$. The action of $b_{\ell m}$ and $b^\dagger_{\ell m}$ on one-particle states $|n\ket_{lm}$ is defined as usual 
\bea\label{bbdagger}
b_{lm}|n\ket_{lm}&=&\sqrt{n}\,|n-1\ket_{lm}\nn
b_{lm}^\dagger|n-1\ket_{lm}&=&\sqrt{n}\,|n\ket_{lm}~.
\eea
In the present case, $PT$ invariance requires $\ell$ to be integer.

We compute the two-point function of this field. Using
\be
\sum_{m=-\ell}^\ell Y_{\ell m}(\Omega)Y_{\ell m}(\Omega')={2\ell+1\over4\pi}\,P_\ell(\cos\th)~,
\ee
where $\th$ is the angle between $\Omega$ and $\Omega'$, we get
\be
\bra 0|\vf(R_1,\Omega_2)\vf(R_2,\Omega_2')|0\ket={1\over4\pi}\sum_{\ell=0}^\infty {R_2^\ell\over R_1^{\ell+1}}\,
P_\ell(\cos\th)~,
\ee
where $R_1>R_2$. Using standard resummation formulas for the Legendre polynomials, we get for the radially ordered product:
\be
\bra0|{\cal R}\left(\vf(x)\vf(x')\right)|0\ket={1\over4\pi|x-x'|}~,
\ee
where $x=(R,\Omega_2)$ and $|x-x'|=\sqrt{R_1^2+R_2^2-2R_1R_2\cos\th}$. The above is of course the Green's function in three dimensions. 

After having defined the two-point function of the elementary boundary field $\vf$, we proceed to define the composite field $\vf(x)^2$. This is subtle because of the obvious divergence when the two elementary operators approach each other. We will use the standard normal ordering prescription. Additionally, in order to compare to the bulk operator, we will renormalize $:\vf(x)^2:$ such that it has zero expectation value on any state, as we will explain presently. 

We start with the normal-ordered operator product  $:\vf(x)\vf(x'):$ and take the limit $x\rightarrow x'$. In the normal-ordered product we find four terms. Let us denote them as follows
\be
:\vf(x)\vf(x'):=\{\vf(x)\vf(x')\} +A_1+A_2
\ee
where
\bea\label{A1A2}
\{\vf(x)\vf(x')\}&=&\sum_{ll'}^\infty\sum_{L=|l-l'|}^{l+l'}\sum_{M=-L}^L\sum_{mm'}
\left(b_{lm}^\dagger b_{l'm'}^\dagger R_1^lR_2^{l'}+{b_{lm}b_{l'm'}\over R_1^{l+1}R_2^{l'+1}}\right){c^{ll'mm'}_{LM}\over\sqrt{(2l+1)(2l'+1)}}\,Y_{LM}(\Omega_2)\nn
A_1&=&\sum_{ll'mm'}{b^\dagger_{lm}b_{l'm'}\over\sqrt{(2l+1)(2l'+1)}}\,{R_1^l\over R_2^{l'+1}}\,Y_{lm}^*(\Omega_2)Y_{l'm'}(\Omega_2')\nn
A_2&=&\sum_{ll'mm'}{b^\dagger_{lm}b_{l'm'}\over\sqrt{(2l+1)(2l'+1)}}\,{R_2^{l'}\over R_1^{l+1}}\,Y_{lm}^*(\Omega_2)Y_{l'm'}(\Omega_2')~.
\eea
Notice that $A_1$ and $A_2$ are normal ordered.  In $\{\vf(x)\vf(x')\}$ we already took the limit $\Omega_2=\Omega_2'$. The limit $R_1=R_2$ is also harmless and sets $A_1=A_2$, hence we take it and define
\be
\{\vf(x)^2\}=\sum_{ll'}^\infty\sum_{L=|l-l'|}^{l+l'}\sum_{M=-L}^L\sum_{mm'}
\left(b_{lm}^\dagger b_{l'm'}^\dagger R^{l+l'}+{b_{lm}b_{l'm'}\over R^{l+l'+2}}\right){c^{ll'mm'}_{LM}\over\sqrt{(2l+1)(2l'+1)}}\,Y_{LM}(\Omega_2)
\ee
Next we discuss the explicit mapping of the bulk and boundary Hilbert spaces. This is easiest to see in the Fock space picture. The bulk operator $\F_{\sm{hol}}$ corresponds to the composite operator $\{\vf(x)^2\}$ which is quadratic in the $b$'s:
\be
b^\dagger_{lm}b^\dagger_{l'm'}|0\ket=|1\ket_{lm}\otimes |1\ket_{l'm'}~.
\ee
Therefore, the correspondence should in principle be between bulk one-particle states  and boundary two-particle states. However, the restriction \eq{coeffs} implies that from the set of boundary multi-particle states, we restrict to the subset that involves particles with the {\it same quantum numbers} $\ell$ and $m$. This amounts to restricting to the following class of correlation functions:
\be
{}_{lm}\bra n'|\cdots |n\ket_{lm}
\ee
where, crucially, $l,m$ are the same on both sides. 



It is on this subspace of the boundary Hilbert space that our boundary operators act. Notice now that on this subspace the expectation value of $A_1$ and $A_2$ may not be zero. In fact, $A_1$ and $A_2$ act are $c$-number operators. We get
\bea
A_1&=&{N\over2L+1}\,{R_1^L\over R_2^{L+1}}\,Y_{LM}^*(\Omega_2)Y_{LM}(\Omega_2')\times {\mbox{id}}\nn
A_2&=&{N\over2L+1}\,{R_2^L\over R_1^{L+1}}\,Y_{LM}(\Omega_2)Y_{LM}^*(\Omega_2')\times {\mbox{id}}~,
\eea
for any state with non-zero occupation number. $N$ is a number operator, and id is the identity operator on the space $L,M$. 



Subtracting $A_1$ and $A_2$ from the definition of the renormalized operator $\{\vf^2(x)\}$ amounts to requiring that the expectation value $\bra n|{\cal O}(x)|n\ket$ be zero not only in the vacuum but also on {\it any} state $|n\ket_{\ell m}$ of the reduced boundary Hilbert space. On the other hand, $\bra m|{\cal O}(x)|n\ket$ as well as any of the higher-point functions of ${\cal O}(x)$ are generically non-zero. An operator with such properties can then be compared with the bulk field $\F_{\sm{hol}}$. This prescription is in accordance with the usual AdS/CFT analysis where the boundary operator is expanded into pure creation and pure annihilation operators (see for example \cite{balakraus,banksetal}). There are no cross-terms in this expansion. This is possible if we make the following identification:

\bea\label{abb}
a_{jlm}&=&\sum_{m_1m_2}c^{\ell\ell m_1m_2}_{lm}\,b_{\ell m_1}b_{\ell m_2}\nn
a_{jlm}^\dagger&=&\sum_{m_1m_2}c^{\ell\ell m_1m_2}_{lm}\,b^\dagger_{\ell m_1}b^\dagger_{\ell m_2}
\eea
that holds when $j=2\ell$. 
Hence we conclude that on this Hilbert space, we indeed have
\be
\F_{\sm{hol}}(x)=\{\vf(x)^2\}
\ee
This is in fact the quantum mechanical counterpart of \eq{squarerelation}! Thus, the absence of cross-terms in that equation automatically ensures that the {\it renormalized} operator $\{\vf(x)^2\}$ is finite. This is in 
accordance with the general expectation that bulk quantities correspond to renormalized boundary quantities. Quantum mechanically, the need to compare bulk and boundary quantities mode by mode corresponds to the statement that the operator $\F$ is evaluated on single-particle states.

Having shown the equivalence of bulk and boundary at the level of operators, in the sub-sector of single-particle states, it now follows that
we can get any correlation function of $\{\vf^2\}$ from a quantum mechanical computation in the bulk. In summary, we have identified the
operator of dimension 1 in the boundary theory
\be
\F_{\sm{hol}}=-{\cal O}(x)=\,\{\vf(x)^2\}
\ee

Having studied the normal vacuum $|0\ket_{\ell m}$ and the excited one-particle states $|n\ket_{\ell m}$, we can ask whether we can compute the
one-point function of ${\cal O}(x)$ in the presence of a source. It is now clear how to do this: this corresponds to computing the expectation
value of ${\cal O}(x)$ in some other vacuum $|\ti 0\ket_{\ell m}$ under which the expectation value of the dual operator is not zero. Which
vacuum this is will depend on the details of the source we add, but it could for example be a coherent state. Whether the expectation value
of ${\cal O}(x)$ is finite in that case will depend on the details of the source. When we study the interacting
case, we will naturally encounter states under which ${\cal O}(x)$ has a non-zero expectation value. The above map between the bulk and boundary Hilbert spaces also allows us to compute 
holographically any higher-point function of ${\cal O}(x)$ from the bulk.

\section{Instantons and Conformal Holography}\label{ich}

In section \ref{classicalhol} we studied the standard holographic picture for the $\phi^4$ bulk theory, which is perturbative in the coupling constant. In section 3 we saw that for the free theory there is another holographic picture, which we called Conformal Holography, where one can compare bulk and boundary quantities exactly. In this section we show how this picture persists in the interacting theory if we expand around the right vacuum.

We will discuss an exact solution of the equations of motion \cite{fubini} with special properties. It is the unique non-trivial solution wich 
preserves 
a large symmetry group. We first briefly recall some of its properties. For more details see \cite{fubini}.

Written in cylinder coordinates \eq{cylindercoords}, the solution of the equation of motion
\be\label{eom4d}
\Box\phi-{\l\over 3!}\,\phi^3=0
\ee
takes the form
\be\label{4dsolution}
\phi(x)=\sqrt{48\over-\l}\,{b\over b^2+R^2}~.
\ee
Our application of this solution will be different from the original one. The idea in \cite{fubini} was to start with a conformal theory in flat space and break conformal invariance spontaneously. The vacuum should preserve a subgroup of the 
conformal group, but in particular it should break dilatations. Such a vacuum is given by the expectation value of the scalar field as above. The equation of motion for the scalar field in flat space preserves the full conformal symmetry in four dimensions, namely $O(5,1)$.


It is easy to see that there are no solutions that are invariant under the full conformal group. One can look for solutions that are invariant under its largest possible subgroup. In flat space, it would be reasonable to
demand that this subgroup is the Poincare group in four dimensions. This would require $\phi$ to be a constant. But in the 
presence of interactions \eq{eom4d}, constant solutions do not exist, and the only Poincare invariant solution is the trivial one! Now instead of demanding invariance under the Poincare group, the next best thing is to demand that the solutions are invariant under a twisted subgroup of $O(5,1)$; namely, it is
possible to find a solution invariant under the six four-dimensional Lorentz rotations and four ``modified translations" which are transformations generated by a linear combination of four-dimensional translations and the special conformal generators. The resulting symmetry group is then $O(4,1)$, which should not be confused with the AdS$_4$  group. Notice that in the original conformal group  $O(5,1)$ inversions, although not a part of its connected part, play a very special role \cite{koller} since they connect translations with special conformal transformations. Those same inversions continue to play an important role also in the smaller symmetry group  $O(4,1)$ since they map ``modified translations" to themselves. In particular, they coincide with the inversions of the AdS background if the scale introduced in \cite{fubini} is identified with the AdS radius, i.e.~$b=L$. We should mention the fact that the original group that one starts with is 
higher-dimensional, $O(5,1)$. This suggests that these theories might have a natural realization in five or six dimensions.

\subsection{Fluctuations Around the Bulk Instanton}

The linearized fluctuations around the instanton solution satisfy the  equation 
\be\label{fluct}
\left(\Box+{24b^2\over(b^2+x^2)^2}\right)\F(r,x)=0~,
\ee
where the $\Box$ is in flat 4d-space metric, and $x^2=r^2+\vec{x}^2$. Crucially, this equation is independent of the bulk coupling $\lambda$. We can follow the standard holographic recipe and solve (\ref{fluct}) perturbatively in $r$ to obtain the 2-point function on the boundary. This will be $SO(3,1)$ invariant since that is the symmetry preserved by the boundary theory. We do this in the appendix.
Notice, however, that it is convenient to go to polar coordinates
$
R=\sqrt{r^2+\vec{x}^2}\,,\,\,\, \Omega_3=(\theta,\phi,\omega)
$. In terms of these, the above equation becomes
\be\label{fluct2}
\left(\pa_R^2+{3\over R}\,\pa_R+{1\over R^2}\,\Delta_{S^3}+{24b^2\over(b^2+R^2)^2}\right)\F(R,\theta,\psi,\vf)=0~.
\ee

In order to solve this equation, it is easiest to rescale the field by an overall factor of $R$, and go back to the Euclidean time coordinate $R/b=e^\tau$. We can then separate variables as before
\be
\label{bulk_solutions}
\F={b\over R}\sum_{jlm}c_{jlm}(\tau)\,{\cal Y}_{jlm}(\Omega_3)~.
\ee
We find the following radial equation for $c_{jlm}$ which is reminiscent of the Schr\"odinger equation for a Posch-Teller potential with energy levels $E_j=\pm(j+1)$ 
\bea\label{radialeq}
c_{jlm}''(\tau)&+&{6\over\cosh^2\tau}\,c_{jlm}(\tau)-E_j^2\,c_{jlm}(\tau)=0\,.
\eea
In order to solve \eq{radialeq}, it is easiest to notice that a further coordinate transformation $z=\tanh\tau$ brings this equation to 
the associated Legendre equation \cite{gradshteyn} with coefficients $\m=\pm(j+1)$, $\n=2$. We give the details in Appendix \ref{interactingapp}.
We find the following general solution
\be\label{solutions}
c_{jlm}(R)=c_{(1)jlm}\,\bar P^{j+1}_2(z)+c_{(2)jlm}\,Q_2^{j+1}(z)~,\,\,\,\, z=\frac{R^2-b^2}{R^2+b^2}\,.
\ee
$Q^\m_\n$ is the usual associated Legendre polynomial, and $\bar P^\m_\n$ is a modified associated Legendre polynomial defined in formula \eq{PP} of Appendix \ref{interactingapp}. 

Let us now discuss the symmetries of the solutions \eq{solutions}. The associated Legendre equation has an obvious symmetry $z\rightarrow-z$. In
equation \eq{radialeq}, this corresponds to the inversion symmetry $R^2/b^2\rightarrow b^2/R^2$. Therefore this equation is invariant under
both $R/b\rightarrow\pm b/R$, which are the two $O(4,1)$ inversion symmetries we discussed in section \ref{geomsetup}. We now notice that the
associated Legendre functions have the following properties:
\bea
\bar P_2^{j+1}(-z)&=&(-1)^{j+1}\,\bar P_2^{j+1}(z)\nn
Q_2^{j+1}(-z)&=&(-)^j\,Q_2^{j+1}(z)~.
\eea
We see that some modes are even under $R^2/b^2\rightarrow b^2/R^2$ and some modes are odd, depending on $j$. Of course, this is as in the case
of an expansion in sines and cosines, where the sines are odd under parity whereas the cosines are even. If we want the solution to have a
definite parity, we keep only half of the modes. Of course, combined $\Omega_2\rightarrow-\Omega_2$, this corresponds
precisely to the classification under parity discussed in section \ref{geomsetup}. This is the interacting version of the two quantization 
schemes discussed in \cite{ais} for the free field theory. The above will become crucial when we discuss quantization of these solutions later.

\subsection{Fluctuations Around the Boundary Instanton}

We now go to the classical boundary theory. As in the bulk case, we expand around the instanton solution
\bea
\vf&\rightarrow& \vf_0+\vf\nn
\vf_0(x)&=&\left(360\over-g\right)^{1/4}\left(b\over b^2+x^2\right)^{1/2}~,
\eea
where we have identified the arbitrary bulk and boundary instanton sizes.
We find the following linear equation for $\vf$
\be\label{bdyfluctuations}
\Box\vf+{15b^2\over(b^2+x^2)^2}\,\vf=0~,
\ee
which also does not depend on the boundary coupling $g$.
We will solve this equation again
in cylinder coordinates restricted to the boundary. We set $R=\sqrt{x^2+y^2+z^2}$. Doing similar manipulations as in the bulk case, we finally
find 
\be\label{bdyinteracting}
\vf(R,\Omega_2)=\sqrt{b\over R}\sum_{lm}\left(\vf_{(1)lm}\,P_{3/2}^{l+\half}(z)+\vf_{(2)lm}\,Q_{3/2}^{l+\half}(z)\right)Y_{lm}(\Omega_2)~,\,\,\,\,z=\frac{R^2-b^2}{R^2+b^2}\,.
\ee
Again, $\vf_{(1)}$ and $\vf_{(2)}$ are arbitrary coefficients.

\subsection{The Bulk-Boundary Correspondence}

In this section we compare the classical solutions of the fluctuation equations around the instanton in AdS$_4$ with the classical fluctuations 
around the boundary instanton. In the free case, comparison between the bulk and the boundary was relatively straightforward. We obtained the
relation
\be
\F_{\sm{hol}}=\vf^2
\ee
as a relation between bulk and boundary modes. In the interacting case, this relation seems to survive, but it is much more involved to establish.
The reason is that dilatations $R\rightarrow\l R$ have been broken on both sides. Therefore, one cannot expect one single bulk mode to correspond
to a mode on the boundary with the same energy. There will be some mixing. Since the analysis is rather involved, we have carried it out 
explicitly only for half of the solutions, namely the $P$'s. We will show this explicitly.

Thus, we are set to compare the bulk solutions to the square of the boundary solutions. To start doing that we write the $P$-Legendre part of the bulk solutions (\ref{bulk_solutions}) in terms of the $\tilde{P}$ modified associated Legendre as 
\be\label{a1}
\F_{\sm{hol}}^+(R,\Omega_2)={b\over R}\sum_{j=0}^\infty\sum_{l=0}^j\sum_{m=-l}^l\ti P_2^{j+1}c_{jlm}\,Y_{lm}~,
\ee
We wish to compare that to the square of 
\be\label{a2}
\vf^+(R,\Omega_2)=\sqrt{b\over R}\sum_{\ell=0}^\infty\sum_{m=-\ell}^\ell P^{\ell+\half}_{3/2}\,\vf_{(1)\ell m}\,Y_{\ell m}~.
\ee
$\ti P_2^{j+1}(z)$ is a linear combination of $P_2^{-(j+1)}(z)$ and $P_2^{-(j+1)}(-z)$. The superscript $+$ reminds us of the fact that we are dealing with half of the modes, and not with the full solution. We will get back to this when we discuss quantization. It is easy to see that the $R=0$ and $R=\infty$ limits agree (this is true for the $Q$'s also). Namely, for $j\geq2$ $\ti P_2^{j+1}$ and $Q_2^{j+1}$ diverge like $\sim 1/R^{j+1}$ at $R=0$, and as $\sim R^{j+1}$ at $R=\infty$. For details, see formula \eq{asympt} in Appendix \ref{interactingapp}.

However, we seek an exact relation. We first consider generic modes $j\geq2$, $\ell\geq2$. The following relation holds
\be\label{interactingsquarerel}
P_{3/2}^{\ell+\half}(z)\,P_{3/2}^{\ell'+\half}(z)=\sum_{j=2}^{\ell+\ell'}d_{j\ell\ell'}\,\ti P_2^{j+1}(z)~,
\ee
where $\ti P$ is defined in \eq{PP} of Appendix \ref{hypersphericalapp}, and $d_{j\ell\ell'}$ are constants which are completely fixed by the
above relation. Using mathematica, we have checked the above relation and computed $d_{j\ell\ell'}$ for values of $\ell,\ell'$ up to 
$\ell+\ell'=10$, and up to $2\ell=16$ in the diagonal case $\ell=\ell'$. Equation \eq{interactingsquarerel}
is a non-trivial statement about associated Legendre functions which has not 
appeared in the literature. An analytic proof of it based on properties of the hypergeometric function will be given in Appendix \ref{thkapp}
by Tom Koornwinder. 
We also notice here that $(P_{3/2})^2$ and $P_2$ have the 
same behavior at $z=\pm1$ (which corresponds to $R=0,\infty$), and their only other singularities in the complex plane are a branch cut at 
${\mbox{Im}}\,z\leq1$ and a pole at $z=\infty$.  Recalling that in the original
coordinates $z=y_4/y_0$, analyzing the singularity structure at $z\rightarrow\infty$ means going beyond $|y_0|>L$, i.e.~it amounts to analyzing the
singularity structure in the embedding Minkowski space-time.

It is easy to understand why in \eq{interactingsquarerel} only terms with $j\leq\ell+\ell'$ appear and not higher. Whereas we have broken 
dilatation invariance $R\rightarrow\l R$, the two fixed points of dilatations, $R=0$ and $R=\infty$, are still preserved by the solutions. In fact, they are related by an inversion. This implies that the leading divergence comes from $j=\ell+\ell'$.





















Having established the relation between the bulk and the boundary for the radial dependence, the important point now is to match the coefficients. We proceed as in the free case, by squaring the full field \eq{a1}. Using \eq{spherharmmult}, we get
\be
(\vf^+(R,\Omega_2))^2=\sum_{\ell,\ell'=0}^\infty\vf^+_\ell\vf^+_{\ell'}~,
\ee
and the bilinears are
\bea
\vf^+_{\ell}\vf^+_{\ell'}&=&{1\over R}\sum_{mm'}P_{3/2}^{\ell+\half}P_{3/2}^{\ell'+\half}\,\vf_{(1)\ell m}\vf_{(1)\ell'm'}\,Y_{\ell m}Y_{\ell'm'}\nn
&=&{1\over R}\sum_{mm'}P_{3/2}^{\ell+\half}P_{3/2}^{\ell'+\half}\vf_{(1)\ell m}\vf_{(1)\ell m'}\sum_{L=|\ell-\ell'|}^{\ell+\ell'}\sum_{M=-L}^L
c_{LM}^{\ell\ell'mm'}Y_{LM}\nn
&=&{1\over R}P_{3/2}^{\ell+\half}P_{3/2}^{\ell'+\half}\sum_{L=|\ell-\ell'|}^{\ell+\ell'}\sum_{M=-L}^Mc^{\ell\ell'}_{LM}Y_{LM}~,
\eea
where $c^{\ell\ell'}_{LM}=\sum_{mm'}\vf_{(1)\ell m}\vf_{(1)\ell'm'}c^{\ell\ell'mm'}_{LM}$. As in the free case, they vanish if $\ell+\ell'+L$ is not even. So, we finally compare the square of the boundary field
\be\label{b1}
(\vf^+(z))^2=\sum_{\ell,\ell'=0}^\infty\vf_\ell^+\vf_{\ell'}^+={1\over R}\sum_{\ell,\ell'=0}^\infty\sum_{j=0}^{\ell+\ell'}\ti P_2^{j+1}d_{j\ell\ell'}\sum_{L=|\ell-\ell'|}^{\ell+\ell'}\sum_{M=-L}^Lc^{\ell\ell'}_{LM}Y_{LM}
\ee
to the bulk result
\be\label{b2}
\F_{\sm{hol}}={1\over R}\sum_{j=0}^\infty\sum_{l=0}^j\sum_{m=-l}^l\ti P_2^{j+1}c_{jlm}Y_{lm}~.
\ee
Using orthogonality of the $Y_{lm}$'s, we can integrate them out. Then we can compare the $P_2$'s mode by mode. We are left with the following relation:
\be
\sum_{\ell,\ell'=0}^\infty  d_{j\ell\ell'}\,c^{\ell\ell'}_{lm}=c_{jlm}
\ee
where the sum runs over $j\leq\ell+\ell'$. Also, from the boundary we get the restriction $L\geq|\ell-\ell'|$, otherwise the left-hand side is zero. Therefore the $c_{jLM}$ should be chosen to be zero if this condition is not satisfied. Thus, we get
\be\label{bulkbdyinter}
\sum_{\ell,\ell'=0}^\infty d_{j\ell\ell'}\sum_{m_1m_2}c^{\ell\ell'm_1m_2}_{lm}\,\vf_{(1)\ell m_1}\vf_{(1)\ell'm_2}=c_{jlm}~.
\ee
As before, let us now restrict to the single-particle states $\ell=\ell'$, leaving for the future the correspondence for the non-diagonal states. We get:
\be\label{bulkbdy}
c_{jlm}=\sum_{\ell=j/2}^\infty d_{j\ell\ell}\sum_{m_1m_2}c^{\ell\ell m_1m_2}_{lm}\,\vf_{(1)\ell m_1}\vf_{(1)\ell m_2}~,
\ee
where as usual $j$ is even. Here, we have used the fact that the boundary sum is restricted by $j\leq2\ell$, as we see from \eq{b1}. This is an 
important relation that we will quantize later. We note here that we can invert \eq{interactingsquarerel} and express the bulk $j$'s as linear
combinations of boundary $\ell$'s. The bulk-boundary relation \eq{bulkbdy} will be inverted as well: pairs of boundary states with energy $\ell$ will be expressed
in linear combinations of the bulk states.

In the left-hand side of equation \eq{bulkbdy}  we have the bulk constraint $l\leq j$, which does not have an obvious parallel on the boundary. 
Apparently, we can only compare modes with total angular momentum less than $j$. However, for the spherically symmetric states $l=0$ we 
can compare all the modes, and it is for these modes that we have full access to the boundary from the above bulk computation. Notice that this 
does not mean that the boundary field is spherically symmetric, but only that the total angular momentum is zero. For the remaining, non-spherical
modes, the bulk result is reproduced from the boundary provided that we restrict $l\leq j$.
It would be interesting to see whether this constraint  can be relaxed. 

Since the bulk-boundary correspondence is most naturally interpreted at the quantum level, we will discuss the meaning of \eq{bulkbdy} after we discuss quantization in the next section.

Of course, we recover the free case when we replace the associated Legendre functions by monomials. This corresponds to setting $d_{j\ell\ell}=\d_{j/2,\ell}$ above. In that case the above sum collapses to just the first term. 

We notice here that a similar comparison for the $Q$'s is more subtle, due to the appearance of logarithms in the boundary computation. In the bulk though we also have logarithmic terms for $j=0,1$. It seems likely that both logarithmic terms can be related to each other. We have not looked at this interesting problem, which we will leave for the future.

\subsection{Quantization}\label{quantization}

Quantization will proceed very close to the free case. It is clear that we will promote the arbitrary coefficients $c_{(1)jlm}$ and $c_{(2)jlm}$ to 
operators that create and annihilate states with quantum numbers $jlm$. However, we need to decide which one is the creation and which one is 
the annihilation operator. In fact, the discussion proceeds as in the free case. The coefficient that multiplies a mode $R^j$ that is regular
at $R=0$ will create a state of energy $j$. Indeed, this corresponds to a mode $e^{j\tau}$ in the time coordinate, which after Wick rotation
becomes $e^{ijt}$ and is the mode associated with a creation operator. This means that the modes that are regular at $R=0$ will couple to
creation operators, and the ones that are regular at $R=\infty$ will couple to annihilation operators.

We will now impose these regularity conditions. Recall that we had the solution
\bea\label{regularsplitting}
\F_{jlm}(R,\Omega_3)&=&{b\over R}\left(c_{(1)jlm}\,\bar P_2^{j+1}(R)+c_{(2)jlm}\,Q_2^{j+1}(R)\right){\cal Y}_{jlm}\nn
&=&\F_{jlm}^+(R,\Omega_3)+\F_{jlm}^-(R,\Omega_3)~,
\eea
where $\F^+$ is the part of the field that is regular at $R=0$, and $\F^-$ the one regular at $R=\infty$. Using the asymptotics of the associated
Legendre functions worked out in Appendix \ref{interactingapp}, we find
\bea\label{regularity}
\F^+_{jlm}(R,\Omega_3)&=&(-)^{j+1}\,{1\over R}\left(\bar P_2^{j+1}(R)+{2\over(j-2)!(j+3)!}\,Q^{j+1}_2(R)\right){\cal Y}_{jlm}^*\nn
\F^-_{jlm}(R,\Omega_3)&=&{1\over R}\left(\bar P_2^{j+1}(R)-{2\over(j-2)!(j+3)!}\,Q^{j+1}_2(R)\right){\cal Y}_{jlm}~.
\eea
By construction, and using the fact that $j+l$ is even, these coefficients satisfy
\bea
\left(\F^+_{jlm}(R,\Omega_3)\right)^\dagger&=&\F^-_{jlm}(R,\Omega_3)\nn
\left(\F^-_{jlm}(R,\Omega_3)\right)^\dagger&=&\F^+_{jlm}(R,\Omega_3)~,
\eea
where Hermitian conjugation is the inversion property discussed earlier: $\f(R)^\dagger={1\over R^2}\,\f(1/R)$. Thus, we finally get
\be
\F(R,\Omega_3)=\sum_{jlm}\left(a_j\,\F^-_{jlm}(R)\,{\cal Y}_{jlm}(\Omega_3)+a_j^\dagger\,\F^+_{jlm}(R)\,{\cal Y}_{jlm}^*(\Omega_3)\right)~,
\ee
Quantization is now standard and proceeds as in the free case. Also, quantization of the boundary theory proceeds analogously, by replacing
the arbitrary coefficients in \eq{bdyinteracting} by operators. In that case, the Hermitian conjuation operation has weight one: $\f(R)^\dagger={1\over R}\,\f(1/R)$.

The key element in comparing the bulk and boundary results is the relation \eq{interactingsquarerel} between the classical solutions, and the
relation \eq{bulkbdy} between the free coefficients which is derived from it. The latter becomes an operator relation. As explained 
earlier, in this paper we quantize the case $\ell=\ell'$ so we have
\bea\label{finaleq}
a_{jlm}&=&\sum_{\ell=j/2}^\infty d_{j\ell\ell}\sum_{m_1m_2}c^{\ell\ell m_1m_2}_{lm}\,b_{\ell m_1}b_{\ell m_2}\nn
a_{jlm}^\dagger&=&\sum_{\ell=j/2}^\infty d_{j\ell\ell}\sum_{m_1m_2}c^{\ell\ell m_1m_2}_{lm}\,b_{\ell m_1}^\dagger b_{\ell m_2}^\dagger~.
\eea
This is the exact modification of the relation \eq{abb} when we include interactions. 

Let us now discuss the physical picture that emerges from the bulk-boundary relation. Clearly, creating a particle in the bulk is like creating a pair of particles on the boundary. More precisely, in the sector $\ell=\ell'$ under consideration, an excitation of energy $j$ in the bulk corresponds to a pair of indistinguishable bosonic excitations with energy $j/2$ on the boundary. Another way to say this is that the bulk Hilbert space is in the tensor product of two copies of the boundary Hilbert space.

As is familiar from AdS/CFT, the energies are given by eigenvalues of the dilatation operator. This picture however gets modified when we include the instanton. The instanton breaks dilatations, and indeed what we find is that every energy level is contributed by different eigenvalues of the dilatation operator. Stated differently, every energy state contributes to an infinite number of dilatation eigenvalues. This is true in the boundary theory as well, and that is in fact the meaning of \eq{finaleq}. A state of energy $j$ in the bulk corresponds to a superposition of boundary states with any $j/2$ and higher. Although the infinite summation range in \eq{finaleq} maybe somewhat unexpected, this is really nothing but the statement that states with energy $\ell$ on the boundary have non-zero intersection with {\it any} state in the bulk with energy up to $2\ell$. 
We should remark here that \eq{interactingsquarerel} and therefore also our final result \eq{finaleq} for the operators can be inverted -- we then
express the boundary states of energy $2\ell$ as superpositions of bulk states of different energies.
This bulk-boundary map is therefore somewhat non-local, if very explicit.

\subsection{Quantization of the Special Modes}

The discussion in the previous section was not entirely complete, for various reasons. In this section we discuss some additional subtleties that appear in the quantization procedure.

In the interacting case there is a special phenomenon concerning normalizability of the modes. As discussed earlier, in Euclidean quantization we associate operators that create particles with positive frequencies to modes that are regular at $R=0$, and operators that annihilate particles with positive frequencies to modes that are regular at $R=\infty$. This procedure therefore clearly distinguishes modes that vanish at $R=0$ and blow up at infinity, from the ones that do the opposite. The relevant linear combinations are listed in \eq{regularity}. For $j=0,1$, however, something special happens, because in that case the $P_2$'s are normalizable (in fact, they go to zero) both at $R=0$ and $R=\infty$. They are self-dual under $R\rightarrow1/R$. The $Q$'s, on the other hand, blow up at both ends. So, {\it any} solution that is regular at one end will be regular at the other end as well. This means that it is impossible to separates modes that create particles from modes that  annihilate 
 them. If we quantize such modes, the field $\F$ will always have a non-zero expectation value in the vacuum. Since we are comparing only the even $j$ modes with the boundary, this means that the $j=0$ mode has a special meaning in the boundary theory and corresponds to coupling the system to a background that sources non-zero expectation values of the composite operator.

There is a mirror of this story on the boundary. The modes $\ell=0,1$ are again special because they are normalizable at both ends. They acquire an expectation value whenever we try to quantize them. Explicitly, we have
\bea
(P_{3/2}^{-1/2}(z))^2&=&a\,z\, P_2^1(z)\nn
(P_{3/2}^{-3/2}(z))^2&=&a_1/z\,P_2^1(z)\,P_2^2(z)\nn
(P_{3/2}^{1/2}(z))^2&=& a_2\,z\,P_2^1(z)+a_3\,(z^2-1)\,Q_2^3(z)\nn
(P_{3/2}^{3/2}(z))^2&=&a_4\,z\left(P_2^{-3}(z)-P_2^{-3}(-z)\right)+a_5\,z\,P_2^1(z)+a_6\,z^2\,Q_2^3(z)
\eea
What we called ``special'' here are only the first two modes, which are indeed written in terms of the special $j=0,1$ bulk modes. The last two modes are special in that their expansion differs from the general expression \eq{interactingsquarerel}, but are otherwise non-normalizable at both ends and should not be quantized. 

Thus, the square of $P_{3/2}^{-1/2}$ corresponds to the $j=0$ bulk mode, as expected. For these modes, the above interpretation applies. On the other hand, the square of $P_{3/2}^{-3/2}$ is itself quadratic in the bulk $j=0$ and $j=1$ modes. This seems to violate the condition $j=2\ell$ and we do not have any explanation for this fact.

The discussion in section \ref{quantization} was also limited for the following reason. We established the quadratic relation \eq{interactingsquarerel} between the bulk and the boundary modes only for the $P$'s. On the right-hand side of \eq{interactingsquarerel}, only $\ti P_2^{j+1}$'s appear, which have a definite symmetry under $R\rightarrow 1/R$ ($z\rightarrow-z$; see the definition \eq{PP}). That makes them non-normalizable both at $R=0$ and at $R=\infty$. This means that they are associated to linear combinations of creation and annihilation operators. In order to get a linear combination that is normalizable at one of the two ends, we need to do as in \eq{regularity}, which also involves the $Q$'s. Keeping only the $P$'s is like doing quantum field theory with only half of the modes, the sines or the cosines. A composite operator built from only half of the modes will again have a non-zero expectation value in the vacuum. Thus, to complete the discussion of quantization, it would be essential to construct the map between $P_{3/2}^{-(\ell+\half)}$ (or $Q_{3/2}^{\ell+\half}$) and $Q_2^{j+1}$.

Finally, we should comment on the possibility of adding classical sources. By this we mean sources on top of the instanton solution which already acts as a source and gives ${\cal O}(x)$ non-zero expectation value. Classical sources are arbitrary, and in particular they can be given by any of the non-normalizable modes of the fluctuation equation. A particularly interesting case is that of $\ell$ half-integer, which is allowed in a classical theory where we violate $PT$ invariance. It seems that much of the bulk-boundary analysis should go through for such modes, and it would be interesting to understand this in detail.

\section{Instanton Decay and de Sitter Space}

Our focus in this paper has been the construction of the exact holographic map between the $\f^4$ theory in the bulk and the boundary $\f^6$ theory. The use of instantons was motivated by the fact that they can be regarded as non-perturbative vacua of the theory. Of course, their physical interest is that they describe tunneling from one solution to another. In this section we explain what may be the relevance of these solutions to tunneling between solutions, leaving a more detailed analysis for the future.

The best way to think of the fluctuation equation in terms of an unstable solution is to notice that it is the equation for a massive scalar field in Euclidean de Sitter space\footnote{After completion of this paper we learned that similar ideas appeared in \cite{loran}.} (in other words, a sphere). Indeed, both in the bulk and boundary theories, the equations of motion for the fluctuations around the instanton, equations \eq{radialeq} and \eq{bdyfluctuations}, are those of a tachyon on a 3- or 4-dimensional sphere, respectively, of the radius of the instanton size (which in the bulk is the scale of AdS). 
Both in the bulk and in the boundary theories, we can have a rough picture of an effective expanding bubble inside which the scalar field behaves 
as a tachyon. We stress the word ``effective'' here, because this is a description of the fluctuations  of the scalar field only.
Let us see how this comes about. We write the $D$-dimensional sphere in the following coordinates
\be
\dd s^2={4b^2\over\Omega^2}\left(\dd R^2+R^2\dd\Omega_{D-1}^2\right)
\ee
where $\Omega=b^2+R^2$. Consider now a massive scalar field on this sphere:
\be
\left(\Box-m^2\right)\f=0~.
\ee
Redefining $\f=(R/b)^{1/2}\,\F$ and $R=b\, e^\tau$, we get the equation of motion:
\be
\F''(\tau)+\left(\Delta_{S^3}-1\right)\F(\tau)+{2-m^2b^2\over\cosh^2\tau}\,\F(\tau)=0
\ee
if $D=4$, and 
\be
\F''(\tau)+\left(\Delta_{S^2}-{1\over4}\right)\F(\tau)+{3-4m^2b^2\over4\cosh^2y}\,\F(\tau)=0
\ee
if $D=3$. Comparing this to the instanton fluctuation equations \eq{fluct2} and \eq{bdyfluctuations}, we get the following tachyonic values of the mass
\bea
m^2&=&-{4\over b^2}~~~\mbox{if}~~D=4\nn
m^2&=&-{3\over b^2}~~~\mbox{if}~~D=3~.
\eea

It is possible to take $\l$ to be positive. In this case, in order to obtain a real solution we also need to analytically continue $b\rightarrow i\,b$.  The solution then looks like:
\be
\f=\sqrt{48\over\l}{b\over b^2-R^2}~.
\ee
This solution has very interesting properties.
Again, there is a boundary solution that is the square root of the above. The fluctuation equations can be studied the same way in this case, and in fact they correspond to the equations of motion of a scalar field in anti-de Sitter space. Now the solutions of the fluctuation equations where $\l>0$ can be obtained from the ones with $\l<0$ by replacing $z$ by $z'=1/z$. Thus, we get the bulk solutions $P_2^{-(j+1)}(z')$ and $Q_2^{j+1}(z')$. The boundary solutions are $P_{3/2}^{\ell+\half}(z')$ and $P_{3/2}^{-(\ell+\half)}(z')$. Again, they are related by the squaring relation.

The seemingly innocent transformation $z'=1/z$ has dramatic consequences. In the case $\l<0$, the singular points $z=\pm 1$ corresponded to the fixed points of dilatations, that is to future and past infinity. The singularity at $z=\infty$ was a point in 5-dimensional Minkowski space outside the AdS hyperboloid. When $\l>0$, the singularity occurs at a real value of $R$, $R=b$, which is where the boundary of the new (effective) anti-de Sitter space is.

\section{Discussion and Outlook}

In this paper we have studied a toy model which, despite its simplicity, seems to capture many interesting physical properties that deserve further study. Apparently, in this model we are able to probe both sides of the AdS/CFT duality semi-classically and get exact agreement. We found that, already in the free theory with no potential, classical bulk fields are given by the square of the boundary fields. This picture persists if we include the potential. The bulk instanton solution is the square of the boundary instanton solution. Then we considered fluctuations around the instanton background, and again found exact agreement for the bulk and boundary $P$-modes. In this case there is a mixing of modes with different conformal dimensions, due to the fact that the instanton breaks dilatation invariance. 

The bulk-boundary correspondence appears to be much more natural upon quantization of the solutions in the bulk and in the boundary theories. Using radial quantization in the bulk, we reproduced the two-point function of the dual composite operator ${\cal O}(x)$ on the boundary. But since we have agreement between the bulk and the boundary mode by mode, we were able to go one step further and identify the composite boundary operator ${\cal O}(x)$ as a normal-ordered product of the elementary boundary field $\vf(x)$, which in turn gets holographically related to the quantized bulk field $\F_{\sm{hol}}$. This was done for any state in the Hilbert space of one-particle states on the boundary. We found that the bulk Hilbert space is  the tensor product of two copies of the boundary Hilbert space, projected onto a diagonal subspace of observables. It is a very interesting open problem to identify the multi-particle states from the bulk, and we gave some indications of how this might work. Having the explicit map, one may expect to be able  in principle to match also higher-point correlation functions in this one-particle subsector.
Moreover, the fact that the quantized bulk field gets identified with the {\it renormalized} boundary composite operator explains the absence of certain mixing terms in the comparison of the classical solutions. We gave the explicit definition of this renormalized field.

We also discussed the physical effect of the instanton background on the fluctuations. We found that the instanton cloaks the fluctuations to find themselves surrounded by an effective de Sitter space. The masses of the fluctuations are tachyonic, which points to the fact that the model describes a decay effect. 
Indeed, when continued back to Lorentzian signature, the solution is time-dependent.
For a full discussion of this issue and its physical implications, the back-reaction of the fluctuations should be taken into account. We note 
though that the solution itself
is an exact solution of Einstein's equations in AdS \cite{nextpaper}.

This toy model seems to be giving us some insight in a special class of theories where both sides of AdS/CFT might be under better control. In our view, this is intimately connected to classical conformal invariance in the bulk and the existence of instanton solutions \cite{nextpaper,sebaspeng}. Bulk conformal invariance guarantees that normalizable modes can reach the boundary. As is well known, massless particles in AdS can reach the boundary in finite time, whereas massive particles are bound to oscillate in the bulk. It is therefore conceivable that there exists an effective theory for the massless modes constructed by a simple rearrangement of the bulk degrees of freedom reduced to the boundary. This is indeed what we find in this paper. The second important point is that the model has instanton solutions. So we are able to find exact vacua of the theory in the interacting case and expand around them.

Let us note here that bulk conformal invariance is a fundamental property of Higher-Spin gauge theories in the frame-like formulation of Vasiliev \cite{Misha}. These theories may also contain instanton-like solutions similar to the the ones considered here \cite{SS}. We believe that the type of holography described in our paper is the appropriate one for the holography of Higher-Spin gauge theories.

A natural extension of our work would be to explicitly compute the 2-point function of the operator ${\cal O}(x)$ in the instanton background and 
compare it with the boundary theory, as we did in the free case. Another interesting continuation of our work would be the 
study of the standard gauge theory instantons in AdS$_4$ and their holography (see e.g. \cite{MaldaMaoz,tassos1}). 
As in the case of standard gauge theory instantons, the Fubini instanton has no back-reaction on the background \cite{nextpaper}. The fluctuations of course do, however the effects of back-reaction can be incorporated systematically in the linearized approximation. At the linearized level, back-reaction introduces a source term on the right-hand side of the fluctuation equation \eq{fluct}. Therefore, the simple form of the holographic dual we have studied in this paper seems to contain information about the solution of the homogeneous equation only. This points to the fact that, in the instanton vacuum, the CFT couples to a source. This situation is then somewhat reminiscent to the discussion in \cite{HM} where similar effects were found. Notice that in the trivial vacuum $\f=0$ discussed in section 3, this effect is absent and at the linearized level the back-reaction can be neglected altogether, its effects appearing only at second order. 
We are currently investigating such issues.

It would be very interesting to see whether a simple relation between classical solutions exists for other operators in the boundary theory -- for the stress-tensor, say. This would already be quite interesting even in the absence of the scalar field. Whereas we do not know the answer to this question, one would expect that, if such a relation exists, it will hold for a restricted class of bulk solutions. In this context it would be interesting to study bulk gravitational instantons.

As is well known, the case of $\lambda<0$ is an approximate solution of ${\cal N}=8$ supergravity in four dimensions \cite{hh}. Here we pointed 
out another form of the solution where the potential is bounded and $\l>0$. Using the results of \cite{yiannis}, this model can be lifted to 
M-theory, where the value of $\l$ is completely determined by the geometric set-up \cite{nextpaper}.


While this paper was being finished, \cite{fssy} appeared, which discusses closely related issues in a different set-up.

\section*{Acknowledgements}

\addcontentsline{toc}{section}{Acknowledgements}

We thank Burkhard Eden, Peng Gao, Massimo Porrati, George Siopsis and Theodore Tomaras for discussions. We also thank each other's institutions for hospitality at various stages of this work. The research of SdH is in part supported by the EC Marie Curie Research Training Network MRTN-CT-2004-512194. The work of A.~C.~P.~is partially supported by the PYTHAGORAS II Research Program of the Greek Ministry of Higher Education.

\appendix

\section{Conventions and Coordinate Systems}\label{conventions}

Here we give the explicit coordinate transformation from Poincare coordinates $(r,\vec{x})$, where the metric takes the form
\be
\dd s^2={\ell^2\over r^2}\left(\dd r^2+\dd\vec x^2\right)~,
\ee
to embedding coordinates $(y_0,\ldots,y_4)$ and cylinder coordinates $(R,\th,\psi,\omega)$. We have
\bea
u&=&r+{\vec{x}^2\over r}\nn
v&=&{\ell^2\over r}\nn
\vec y&=&{\ell\over r}\,\vec x
\eea
and
\bea
x&=&R\,\sin\th\sin\psi\sin\omega\nn
y&=&R\,\sin\th\sin\psi\cos\omega\nn
z&=&R\,\sin\th\cos\psi\nn
r&=&R\,\cos\th~.
\eea
The range of the angles is $0\leq\th\leq\pi/2$, $0\leq\psi\leq\pi$, $0\leq\omega\leq2\pi$.

\section{Hyperspherical Harmonics and Holography in Cylinder Coordinates}\label{hypersphericalapp}

The hyperspherical harmonics ${\cal Y}_{jlm}$'s satisfy
\bea
\Delta_{S^3}{\cal Y}_{jlm}&=&-j(j+2){\cal Y}_{jlm}~\,,\,\,\,\, j=0,1,2...\\
\label{hypersph1}
\int_{S^3} d\Omega_2 {\cal Y}^*_{jlm}(\Omega_3) {\cal Y}_{j'l'm'}(\Omega_3)&=&\delta_{jj'}\delta_{ll'}\delta_{mm'} \\
\label{hypersph2}
\sum_{m=-l}^{l} {\cal Y}^*_{jlm}(\Omega_3){\cal Y}_{jlm}(\Omega_3') &=&{2l+1\over4\pi}\,N_{jl}^2\,(\sin\theta\sin\theta')^l\,\times\\
&\times&C_{j-l}^{l+1}(\cos\theta)\,C_{j-l}^{l+1}(\cos\theta')\,P_l(\cos(\phi-\phi'))\\
{\cal Y}_{jlm}(\Omega_3) &=&N_{jl} (\sin\theta)^lC_{j-l}^{l+1}(\cos\theta) Y_{lm}(\Omega_2)\label{hyp}\\
N_{jl} &=& 2^l\Gamma(l+1)\left(\frac{2(j+1)}{\pi}\right)^{\frac{1}{2}} \left(\frac{\Gamma(j-l+1)}{\Gamma(j+l+2)}\right)^{\frac{1}{2}}
\eea
where $Y_{lm}(\Omega_2)$ are the standard two-dimensional spherical harmonics, $C_m^n(t)$ are Gegenbauer polynomials and $P_l(t)$ are Legendre functions. 

When reduced to the boundary, the hyperspherical harmonics reduce to the spherical harmonics as in \eq{jlmreduction}. The proportionality 
constants are given by
\be\label{ajl}
a_{jl}=N_{jl}\,{({j+l\over2}+1)!\over({j-l\over2})!}\,(-)^{j-l\over2}
\ee
if $j+l$ is even, and zero otherwise.

Recall the multiplication property of spherical harmonics
\be\label{spherharmmult}
Y_{j_1m_1}(\Omega_2)Y_{j_2m_2}(\Omega_2)=\sum_{L=|j_1-j_2|}^{j_1+j_2}\sum_{M=-L}^Mc^{j_1j_2m_1m_2}_{LM}\, Y_{LM}(\Omega_2)~.
\ee
The $c^{ll'mm'}_{LM}$'s are given by \cite{messiah}
\bea\label{coeffs2}
c^{ll'mm'}_{LM}&=&(-)^M\sqrt{(2l+1)(2l'+1)(2L+1)\over4\pi}\left({l\,l'L\atop000}\right) \left({l~~l'~~~L\atop mm'-M}\right)\nn
\left({abc\atop000}\right)&=&(-)^p\sqrt{\Delta(abc)}\,{p!\over(p-a)!(p-b)!(p-c)!}\nn
\Delta(abc)&=&{(a+b-c)!(b+c-a)!(c+a-b)!\over(a+b+c+1)!}\nn
p&=&{l+l'+L\over2}~.
\eea
The coefficients vanish unless $j_1+j_2+L$ is even, a property that will be crucial when comparing with the bulk.

Expanding in hyperspherical harmonics captures the normalizable bulk modes. In AdS/CFT one also considers non-normalizable modes, which 
correspond to classical sources on the boundary. We now provide the relevant analysis in cylinder coordinates (for the standard Poincare analysis, see section \ref{standardsol}). This analysis is also relevant to
the theory with an operator of dimension 2. The idea is to solve the second
order differential equation for the holographic coordinate $\th$. To that end we write the Laplacian on the three-sphere as a circle fibration over the
the $S^2$. Thus we solve the differential equation
\be
{1\over\sin^2\th}\,\pa_\th\left(\sin^2\th\,\pa_\th\,\F_{j,\ell}\right)-{\ell(\ell+1)\over\sin^2\th}\,\F_{j,\ell}=-j(j+2)\,\F_{j,\ell}~.
\ee
Rescaling $\F$, and performing a coordinate transformation $z=\cos\th$, we get the associated Legendre equation \eq{alf} in Appendix \ref{interactingapp}
with $\n=j+\half$, $\m=\ell+\half$. The general solution is
\be
\F_{j,\ell}(\th)={1\over\sqrt{\sin\th}}\left(d_{(1)}\,P_{j+\half}^{\ell+\half}(\cos\th) +d_{(2)}\,P_{j+\half}^{-(\ell+\half)}(\cos\th)\right)~.
\ee
As expected, the behavior at $\th\rightarrow\pi/2$ distinguishes two cases: $j+\ell$ even or odd. In the even case, the asymptotics as $\th\rightarrow\pi/2$ is
\bea
P^{\ell+\half}_{j+\half}(\cos\th)&\sim&\cos\th\nn
P^{-(\ell+\half)}_{j+\half}(\cos\th)&\sim&1~,
\eea
whereas in the odd case these two get interchanged. Taking into account the fact that $\F$ has been rescaled by $r=R\cos\th$, the mode that
goes asymptotically to a constant corresponds to the expectation value of an operator of dimension 1, and the one that vanishes as $\cos\th$ is the 
operator of dimension 2. Alternately, we can think of a single theory with an operator and a source.

Imposing regularity at $\th=0$ rules out the $P^{\ell+\half}_{j+\half}$ modes in both cases. We are left with the mode 
\be
{1\over\sqrt{\sin\th}}\,P^{-(\ell+\half)}_{j+\half}(\cos\th)=\sqrt{2\over\pi}\,{2^\ell \ell!(j-\ell)!\over(j+\ell+1)!}\,(\sin\th)^\ell \,C_{j-\ell}^{\ell+1}(\cos\th)~,
\ee
and we of course recover the hyperspherical harmonics \eq{hyp}. In conclusion, if we are in the theory with the operator of dimension 1, then $j+\ell$ has
to be even.

\section{Solutions in the Interacting Case}\label{interactingapp}

We first define two sets of associated Legendre polynomials used in the main text:
\bea\label{PP}
\bar P^{j+1}_2(z)&=&P^{-(j+1)}_2(z)+(-1)^{j+1}P^{-(j+1)}_2(-z)\nn
\ti P^{j+1}_2(z)&=&P^{-(j+1)}_2(z)+(-1)^jP^{-(j+1)}_2(-z)
\eea
for $j\geq2$. For $j=0,1$, $\bar P^{j+1}_2=\ti P_2^{j+1}=P^{j+1}_2$.

Next we find the solutions of \eq{radialeq}. Performing a coordinate transformation $z=\tanh \tau/b$, we can bring the radial equation to the 
following form
\be\label{alf}
{\dd\over\dd z}(1-z^2){\dd\over\dd z}\vf+\left(\nu(\n+1)-{\m^2\over1-z^2}\right)\vf=0
\ee
where $\n=2$ and $\m=\pm(j+1)$. This is the associated Legendre equation with solutions $P_\n^\m$ and $Q_\n^\m$. Since $\n$ and $\m$ are 
integral, we have to distinguish the cases $j=0,1$ and $j>1$ \cite{gradshteyn}. \\
\\
\underline{\bf $j=0,1$}\\
\\
We discuss this case first. We have the two independent solutions
\be\label{firstset}
P_2^{j+1}(z)~,~Q^{j+1}(z)~.
\ee
In particular,
\bea
P_2^1(R)&=&-6bR\,{R^2-b^2\over(R^2+b^2)^2}\nn
P_2^2(R)&=&{12b^2R^2\over(R^2+b^2)^2}~,
\eea
which was used in the main text. 

It is easy to see that under parity symmetry $z\leftrightarrow-z$, we have
\bea\label{antisym}
P_2^{j+1}(-z)&=&(-)^{j+1}P_2^{j+1}(z)\nn
Q_2^{j+1}(-z)&=&(-)^jQ_2^{j+1}(z)
\eea
for $j=0,1$.\\
\\
\underline{\bf $j\geq2$}\\
\\
The independent set of solutions is now
\be\label{secondset}
P_2^{-(j+1)}(z)~,~Q^{j+1}(z)~.
\ee
However, $P^{-\m}_\n(z)$ does not have any definite symmetry under $z\rightarrow-z$. We can construct a solution with the desired symmetry by
taking a linear combination of $P^{-\m}_\n(z)$ and $P^{-\m}_\n(-z)$. Thus, we replace the $P$'s by either of the two sets $\bar P^{j+1}$,
$\ti P^{j+1}_2$, defined in \eq{PP}. Which one one decides to use is a matter of convention. In the rest of the appendix we will consider
$\bar P$.
Thus, our set of independent solutions is $\ti P^{j+1}_2(z)$, $Q^{j+1}_2(z)$ {\it for any $j$}. By construction, they satisfy
\bea\label{asymP}
\bar P^{j+1}_2(-z)&=&(-)^{j+1}\,\bar P^{j+1}_2(z)\nn
Q^{j+1}_2(-z)&=&(-)^j\,Q^{j+1}_2(z)~.
\eea

\subsection*{Asymptotics of the Solutions}

Here we list the asymptotics of the Legendre functions, which we used in deriving the regularity conditions \eq{regularity}
\bea\label{asympt}
P_2^{-(j+1)}(R)&=&{j(j-1)\over(j+3)!}{1\over R^{j+1}}~,~~~~~~~~~~~~~~R\rightarrow0\nn
P_2^{-(j+1)}(R)&=&{1\over(j+1)!}{1\over R^{j+1}}~,~~~~~~~~~~~~~~\,R\rightarrow\infty\nn
\bar P_2^{j+1}(R)&=&{j(j-1)\over(j+3)!}{1\over R^{j+1}}~,~~~~~~~~~~~~~~R\rightarrow0\nn
\bar P_2^{j+1}(R)&=&(-)^{j+1}\,{j(j-1)\over(j+3)!}\,R^{j+1}~,~~~~~R\rightarrow\infty\nn
Q_2^{j+1}(R)&=&-{j!\over2}{1\over R^{j+1}}~,~~~~~~~~~~~~~~~~~~~~R\rightarrow0\nn
Q_2^{j+1}(R)&=&(-)^{j+1}\,{j!\over2}\,R^{j+1}~,~~~~~~~~~~~~~\,R\rightarrow\infty~,
\eea
for any $j\geq2$, and as usual we defined $P^\m_\n(R)\equiv P^\m_\n(z)$ with $z={R^2-1\over R^2+1}$. Of course, taking into account the rescaling
of $\F$ with an overall $1/R$, we recover exactly the behavior in the free case, \eq{PhiLaplace}.

It is now easy to see why we needed to introduce the $\ti P$'s. $P$ has the same behavior at zero and at infinity: it falls off with the same
power at both ends, so it is regular at zero and it vanishes at infinity. $Q$, on the other hand, falls off with different powers, and in fact it
diverges both at zero and at infinity. Now the most general solution of the equation consists of two modes. If
we want to separate the mode that is regular at zero and diverges at infinity from the one that diverges at zero and is regular at infinity, we
need to replace $P$ by $\ti P$. Now both modes diverge at both ends, and the linear combinations \eq{regularity} have the desired regularity
properties.

In the special cases $j=0,1$, since $P$ by itself was regular at both ends but $Q$ diverges, there is no way to construct a general solution
that is regular at either end but to drop $Q$ for $j=0,1$. The asymptotics of $P$ at $R\rightarrow0$ is
\bea
P_2^1(R)&=&a_0R\nn
P_2^2(R)&=&a_1R^2
\eea
and at $R\rightarrow\infty$,
\bea
P_2^1(R)&=&-{a_0\over R}\nn
P_2^2(R)&=&{a_1\over R^2}~,
\eea
with $a_0=6$, $a_1=12$.

Translating the above for the regularized modes $\F^+_{jlm}(R)$ and $\F^-_{jlm}(R)$ of \eq{regularsplitting}, we find the asymptotic behavior at 
$R=0$
\bea
\F^+_{jlm}(R)&=&{2\over(j+1)!}\,R^j\nn
\F^-_{jlm}(R)&=&(-)^{j+1}\,{2j(j-1)\over(j+3)!}{1\over R^{j+2}}
\eea
At $R=\infty$, we have
\bea
\F^+_{jlm}(R)&=&(-)^{j+1}\,{2j(j-1)\over(j+3)!}\,R^j\nn
\F^-_{jlm}(R)&=&{2\over(j+1)!}{1\over R^{j+2}}~.
\eea
Of course, the parity properties imply that their boundary values are related.

\section{Fluctuation Equation in Poincare Coordinates}\label{standardsol}

In this appendix we solve the fluctuation equation in the standard Poincare form, which we use to
prove the claim that if we impose the regularity condition $\ti\F(r=\infty,\vec x)=0$ and at the same time set $\ti\F(\vec x)=0$, then also $\ti\F_1(\vec x)=0$, made in section 2.2. For notational simplicity we drop the tildes.

The general solution of the fluctuation equation in Poincare coordinates \eq{lineom} is obtained as usual \cite{SSK}:
\be
\F(r,x)=\F_{(0)}(x)+r\,\F_{(1)}(x)+r^2\,\F_{(2)}(x)+\ldots~.
\ee
We get:
\bea
&&2\F_{(2)}(x)+\Box \F_{(0)}(x)+{24b^2\over(b^2+x^2)^2}\,\F_{(0)}(x)=0\nn
&&6\F_{(3)}(x)+\Box \F_{(1)}(x)+{24b^2\over(b^2+x^2)^2}\,\F_{(1)}(x)=0\nn
&&12\F_{(4)}(x)+\Box \F_{(2)}(x)+{24b^2\over(b^2+x^2)^2}\,\F_{(2)}(x)-{2\over(b^2+x^2)^3}\,\F_{(0)}(x)=0~.
\eea
These equations should be viewed as determining $\F_{(2)}$ and $\F_{(3)}$, etc., once $\F_{(0)}$ and $\F_{(1)}$ are provided, as usual. So $\F_{(0)}$ and $\F_{(1)}$ have the interpretations as source/operator in the instanton
background. Notice that, if we set $\F_{(0)}=0$, we get an expansion in odd powers of $r$, whereas the expansion is even if $\F_{(1)}=0$. In fact, we can solve the equations to all orders:
\be
\F(r,x)=\sum_{n=0}^\infty r^n\,\F_{(n)}(x)~.
\ee
We get the following solution:
\be
(n+1)(n+2)\F_{(n+2)}(x)+\Box \F_{(n)}(x)+{24b^2\over(b^2+x^2)^2}\sum_{mk}{(-1)^mm!\over k!(m-k)!}{2^k\over(b^2+x^2)^{2m-k}}\,\F_{(n+2k-4m)}=0
\ee
where the sum runs over $k=0,\ldots,m$, and $n+2k-4m\geq0$. This simplifies to:
\be
\F_{(n+2)}(x)=-{1\over(n+1)(n+2)}\left[\Box \F_{(n)}(x)+24b^2\sum_p{2^{-p}c_p\over(b^2+x^2)^{p+2}}\,\F_{(n-2p)}(x)\right]
\ee
with $c_p=\sum_{m\leq p}{(-1)^mm!2^{2m}\over(2m-p)!(p-m)!}$, and the sum again is such that $n-2p\geq0$. 

In the absence of the instanton, we can in fact resum the series. We get:
\be
\F(r,x)=\cos(r\sqrt{\Box})\F_{(0)}(x)+{1\over\sqrt{\Box}}\,\sin(r\sqrt{\Box})\F_{(1)}(x)~,
\ee
which is the Fourier transform of the usual $\sinh$ and $\cosh$ solutions that correspond to Dirichlet and Neumann boundary conditions.

The proof of the claim in section 2.2 now straightforwardly follows from the decoupling of the even and odd $r$-powers. To analyze the behavior at $r=\infty$ now, it is convenient to use the Euclidean time coordinate $\tau$ in (\ref{radialeq}). At $\tau=\infty$, we can neglect the potential of the fluctuation equation, and we get a free wave equation with solution: $\F=A\,e^{j\tau}+B\,e^{-(j+2)\tau}$. Regularity imposes $A=0$. Now going back to the coordinate $r$, we write $e^{\tau/b}=r/b\,\sqrt{1+\vec x^2/r^2}$. Thus, for odd values of $j$, the expansion will contain both even and odd powers of $r$. But this cannot happen if we set $\F_{(0)}=0$, therefore setting $\F_{(0)}=0$ also requires $B=0$, hence $\F=0$.

\section{Proof of formula \eqref{interactingsquarerel} (by Tom Koornwinder)}\label{thkapp}

{\large(University of Amsterdam, {\tt thk@science.uva.nl})}\\[\medskipamount]
\noindent
Recall the
Pochhammer symbol $(a)_k:=a(a+1)\ldots(a+k-1)=\G(a+k)/\G(a)$
($k$ nonnegative integer), the Gauss hypergeometric series
\[
\hyp21{a,b}cz:=\sum_{k=0}^\iy \frac{(a)_k\,(b)_k}{(c)_k\,k!}\,z^k\qquad
(|z|<1),
\]
its transformation formula
\begin{equation}
\hyp21{a,b}cz=(1-z)^{c-a-b}\hyp21{c-a,c-b}cz,
\label{10}
\end{equation}
and Gegenbauer polynomials expressed in terms of hypergeometric series:
\begin{equation}
C_n^\l(x)=\frac{(2\l)_n}{n!}\,
\hyp21{-n,n+2\l}{\l+\thalf}{\thalf(1-x)}.
\label{3}
\end{equation}
See formulas 9.100, 9.131-1 and 8.932-1 in \cite{gradshteyn}.

We will express the associated Legendre functions occurring on both sides
of \eqref{interactingsquarerel} in terms of Gegenbauer polynomials.
First we deal with the
associated Legendre function $P_{3/2}^{l+\half}(z)$ on the \LHS\ of
\eqref{interactingsquarerel}.
Formula 8.723-1 in \cite{gradshteyn}
writes $P_\nu^\mu(z)$ for large positive $z$
as a linear combination of
two hypergeometric functions of argument
$\thalf-\thalf z(z^2-1)^{-\half}$, but the second term vanishes if
$\mu-\nu-1$ is a nonnegative integer because of a factor $1/\G(\nu-\mu+1)$
in its coefficient. Thus we obtain for integer $l\ge2$ and for large
positive $z$:
\begin{align}
&P_\nu^{\nu+l-1}(z)
\nonumber\\
&=\frac{\G(-\thalf-\nu)}{(2\pi)^\half\G(-2\nu-l+1)}\,
\frac{(z-\sqrt{z^2-1})^{\nu+\half}}{(z^2-1)^{\frac14}}\,
\hyp21{\nu+l-\thalf,-\nu-l+\tfrac32}{\nu+l+\thalf}
{\thalf\left(1-\frac z{\sqrt{z^2-1}}\right)}
\nonumber\\
&=\frac{\G(-\thalf-\nu)}{2^{1+\nu}\pi^\half\G(-2\nu-l-1)}\,
(z^2-1)^{-\half\nu-\half}\,
\hyp21{-l+2,2\nu+l}{\tfrac32+\nu}
{\thalf\left(1-\frac z{\sqrt{z^2-1}}\right)}
\nonumber\\
&=\frac{2^{\nu+1}(-1)^l (l-2)!}{\G(-\nu)}\,(z^2-1)^{-\half\nu-\half}\,
C_{l-2}^{\nu+1}\left(\frac z{\sqrt{z^2-1}}\right).
\nonumber
\end{align}
Here we used \eqref{10} in the second equality, and \eqref{3} together
with the duplication formula of the gamma function in the
third equality. Now put $\nu:=3/2$:
\begin{equation}
P_{3/2}^{l+\half}(z)=
3\sqrt{\frac2\pi}\,(-1)^l(l-2)!\,(z^2-1)^{-\frac54}\,
C_{l-2}^{5/2}\left(\frac z{\sqrt{z^2-1}}\right).
\label{12}
\end{equation}

Next we deal with the function
$\tilde P_2^{j+1}(z)$ on the \RHS\ of \eqref{interactingsquarerel}.
Consider 8.723-1 in \cite{gradshteyn}
for $\nu=2$ and $\mu=j-1$ ($j$ integer $\ge2$)
and apply \eqref{10} to both hypergeometric series occurring on the \RHS.
We obtain for large positive $z$:
\begin{align}
P_2^{-j-1}(z)=&-\,\frac{(z^2-1)^{-3/2}}{15\,(j-2)!}\,
\hyp21{-j+2,j+4}{7/2}{\thalf\left(1-\frac z{\sqrt{z^2-1}}\right)}
\nonumber\\
&+\frac{3\,(z^2-1)}{(j+3)!}\,
\hyp21{-j-3,j-1}{-3/2}{\thalf\left(1-\frac z{\sqrt{z^2-1}}\right)}.
\label{11}
\end{align}
By \eqref{3} the first hypergeometric function in \eqref{11} is proportional
to the Gegenbauer polynomial $C_{j-2}^3\bigl(z(z^2-1)^{-\half}\bigr)$
and the second to $C_{j+3}^{-2}\bigl(z(z^2-1)^{-\half}\bigr)$ (except
that the normalization in \eqref{3} would introduce an artificial singularity
in this second case). We know that $C_n^\l(x)$ is even or odd in $x$, of the
same parity as the degree $n$. Hence the first term on the \RHS\ of
\eqref{11} has the same parity in $z$ as $j-2$, and the second term has the
same parity in $z$ as $j+3$. Hence, by \eqref{PP} and by \eqref{3} we
obtain from \eqref{11} for integer $j\ge2$:
\begin{align}
\tilde P_2^{j+1}(z)=&-\,\frac{16\,(z^2-1)^{-3/2}}{(j+3)!}\,
C_{j-2}^3\left(\frac z{\sqrt{z^2-1}}\right),
\label{13}
\\
\overline P_2^{j+1}(z)=&\frac{6\,(z^2-1)}{(j+3)!}\,
\hyp21{-j-3,j-1}{-3/2}{\thalf\left(1-\frac z{\sqrt{z^2-1}}\right)}.
\end{align}

Now substitute \eqref{12} and \eqref{13} in
the conjectured identity \eqref{interactingsquarerel}.
Then we see that this identity indeed holds with summation only over $j=2,\ldots,l+l'$
with $l+l'-j$ even and with the coefficients $d_{j,l,l'}$ uniquely
determined by the expansion
\begin{equation}
\tfrac98(-1)^{l+l'-1}(l-2)!\,(l'-2)!\,(x^2-1)\,
C_{l-2}^{5/2}(x)\,C_{l'-2}^{5/2}(x)=
\sum_{\bisub{j=2,\ldots,l+l'}{l+l'-j\;{\rm even}}}
\frac{d_{j,l,l'}}{(j+3)!}\,C_{j-2}^3(x).
\end{equation}
Indeed, on the left we have a polynomial of degree $l+l'-2$ in $x$,
even or odd according to whether $l+l'$ is even or odd.

In principle, the coefficients $d_{j,l,l'}$ can be computed from the
known analytic expressions for the coefficients in Dougall's
linearization formula
\begin{equation}
C_m^\l(x) C_l^\l(x)=\sum_{k=0}^{\min(m,l)}a_\l(k,l,m)C_{l+m-2k}^\l(x)
\end{equation}
(see \cite[Theorem 6.8.2]{AAR}),
in Gegenbauer's connection formula
\begin{equation}
C_n^\l(x)=\sum_{k=0}^{[\half n]} b_{\l,\mu}(n,k) C_{n-2k}^\mu(x)
\end{equation}
(see \cite[Theorem 7.1.4']{AAR}), and in the recurrence relation
\begin{equation}
x^2 C_n^\mu(x)=c_\mu(n,n+2)\,C_{n+2}^\mu(x)+c_\mu(n,n)\,C_n^\mu(x)+
c_\mu(n,n-2)\,C_{n-2}^\mu(x)
\end{equation}
(iterate the recurrence relation \cite[(6.4.16)]{AAR}).
However, it seems improbable that the resulting expression for the
coefficients $d_{j,l,l'}$ can be reduced to an analytic expression not involving a sum.

\end{document}